\documentclass[
superscriptaddress,
notitlepage,
showpacs,
amsmath,amssymb,
aps,
longbibliography,
]{revtex4-2}

\usepackage{psfrag,graphicx,epsfig,color}
\usepackage[justification=raggedright, singlelinecheck=false]{caption}
\usepackage{dcolumn}
\usepackage{bm}
\usepackage{subfig}
\usepackage{booktabs}
\usepackage{multirow}
\usepackage{xcolor}
\usepackage{float}
\usepackage{nicefrac}
\usepackage{natbib}
\usepackage{comment}
\usepackage[mathlines]{lineno} 
\usepackage{tabularx}

\newcommand{\chim}{\chi/\langle\chi\rangle}
\newcommand{\avg}[1]{\left \langle #1 \right \rangle}

\newcommand{\ghat}{\hat{\mathbf{g}}}
\newcommand{\omhat}{\hat{\boldsymbol{\omega}}}
\newcommand{\re}{Re_\lambda}

\def\ww {{\boldsymbol{\omega}}}
\def\uu {{\mathbf{u}}}

\def\ff {{\mathbf{f}}}

\def\bg {{\mathbf{g}}}
\def\BG {{\mathbf{G}}}

\begin{document}

\title{Scalar gradient structure and dynamics in turbulent mixing \\ 
at high Reynolds and Schmidt numbers}

\author{Ruqaiya Islam Mishi}
\affiliation{Department of Mechanical and Aerospace Engineering, Texas Tech University, Lubbock, TX 79409}

\author{Dhawal Buaria }
\email[]{dhawal.buaria@ttu.edu}
\affiliation{Department of Mechanical and Aerospace Engineering, Texas Tech University, Lubbock, TX 79409}
\affiliation{Max Planck Institute for Dynamics and Self-Organization, 37077 G\"ottingen, Germany}

\begin{abstract}

How well turbulence mixes a scalar $\theta$ 
is ultimately governed by the scalar dissipation rate 
$\chi = 2D |\nabla\theta|^2$,
where $D$ is the scalar diffusivity, making  
scalar gradients central to turbulent mixing.
We study the structure and amplification of these gradients for the 
canonical problem of passive scalar driven by a
uniform mean-gradient in stationary isotropic turbulence, using well-resolved
direct numerical simulations (DNS) at grid resolutions up to $8192^3$.
The Taylor-scale Reynolds number $\re$ spans $140-1000$, and the 
Schmidt number $Sc\equiv\nu/D$ spans $1-512$, where $\nu$ is the kinematic
viscosity.  We analyze joint statistical correlations
of velocity and scalar gradients that underlie scalar-gradient amplification. 
Unconditional statistics reaffirm earlier observations that production
of scalar dissipation is dominated by nonlinear amplification of scalar gradients by 
strain-rate. Scalar gradients preferentially align with the most compressive
strain eigenvector and remain orthogonal to vorticity,
with  both trends virtually independent of $\re$ and $Sc$. 
Conditional statistics reveal that this organization becomes dramatically
enhanced in regions of intense scalar dissipation: scalar gradient
becomes near-perfectly aligned with the most compressive strain eigendirection
and orthogonal to the other strain eigendirections and to vorticity. 
This geometry, also supported by visualizations, suggests that intense scalar
dissipation is organized in sheet-like structures  
formed in shear layers
between vortex tubes, where intense strain also generally resides. 
However, the effective strain acting along intense scalar gradients is
comparatively much weaker, indicating intense scalar dissipation arises primarily from 
optimal alignments rather than intense strain alone.  
Molecular diffusion arrests intense scalar-gradient events primarily 
by redistributing scalar-gradient variance away from intense structures. 
The contribution from imposed mean-gradient is negligible, 
but still imprints anisotropy directly onto smallest
scales via the strain field. 
The conditional statistics broadly exhibit 
progressively weaker $Sc$-dependence as $\re$ increases,
demonstrating convergence towards an asymptotic state governing the 
local structure and dynamics of intense scalar dissipation.

\end{abstract}

\maketitle

\section{Introduction}
\label{sec:intro}

Turbulent flows are remarkably efficient at mixing scalars, such as heat
(temperature) or substance concentration, in a wide range of natural and 
engineering processes, including atmospheric and oceanic transport, 
cloud physics, combustion, and industrial mixing 
\cite{thorpe2005turbulent, wyngaard2010turbulence, BMSS10, Pitsch2000, Warhaft2000}. 
This dramatic enhancement arises from the ability of turbulence
to continuously deform scalar fluctuations, transferring its variance
from large injection scales to progressively smaller scales -- analogous to the
cascade of kinetic energy -- until molecular diffusion becomes sufficiently effective
to dissipate the smallest scalar structures, thereby completing the mixing
process at the molecular level.
Like the turbulent energy cascade, scalar mixing is also strongly
intermittent, with the dissipation of scalar variance concentrated
in rare and localized extreme events rather than distributed
uniformly throughout the flow. In fact, scalar intermittency 
is known to be stronger than that of the velocity  
\cite{Sreeni97, Shraiman2000, Warhaft2000, BS2022}, 
reflecting the extraordinary efficiency 
with which turbulence generates intense scalar gradients.
Such extreme events play a critical role in many situations.
For instance, in combustion, sharp gradients of fuel 
and temperature can influence
local reaction rates, ignition, flame extinction, 
soot and NOX formation \cite{ Pitsch2000, Sreeni04, Attili2019}. 
In atmospheric and oceanic turbulence, 
strong gradients of temperature,
humidity, or salinity govern entrainment across inversion layers, 
droplet and cloud formation, and the transport of heat, moisture, and dissolved
species over a broad range of scales 
\cite{thorpe2005turbulent, wyngaard2010turbulence}. 
Consequently, understanding the dynamics and statistics of extreme scalar gradients 
is essential both for advancing fundamental 
theories of turbulent mixing and for developing accurate predictive models.

The local rate at which scalar fluctuations are destroyed is quantified by the
scalar dissipation rate:
\begin{align}
 \chi = 2 D g_i g_i  \ , \qquad  g_i = \partial \theta /\partial x_i   
\end{align}
where $D$ is the scalar diffusivity, $\theta$ is the fluctuating
scalar field with $\mathbf{g}$ being its gradient. 
Evidently, the formation of extreme scalar dissipation events is
fundamentally a problem of scalar-gradient amplification. 
For the canonical problem of passive scalar evolving in presence
of a uniform mean gradient $\mathbf{G}$, 
the transport equation for scalar gradient is 
\footnote{Obtained by taking the gradient
of scalar transport equation given later in Eq.~\eqref{eq:theta}}
\begin{align}
\frac{D g_i}{D t} = -A_{ji} g_j - A_{ji} G_j  + D \nabla^2 g_i \ , 
\label{eq:gi}
\end{align}
where $A_{ij} = \partial u_i /\partial x_j$ is the velocity gradient tensor. 
This equation shows that scalar-gradient amplification is governed 
by the interaction of local velocity gradients with both the 
fluctuating scalar gradient and the imposed mean-gradient, 
while diffusion acts to smooth gradients and oppose their growth.
The relative importance of these processes, and mixing dynamics in general,
is controlled by the Reynolds number $Re$, which determines the range and intensity
of velocity-gradient fluctuations \cite{Sreeni97, buaria_jfmp},
and the Schmidt number $Sc \equiv \nu/D$, where $\nu$ is the kinematic
viscosity of the fluid, which sets relative effectiveness of 
scalar diffusion and the separation between smallest
scales in the velocity and scalar fields \cite{Batchelor1959}. 

Several studies have sought to understand how turbulence 
amplifies scalar gradients by examining their geometric 
relationship with the local velocity gradients, 
particularly via the eigenframe of the strain-rate tensor, 
the symmetric part of the velocity gradient tensor
\cite{Ashurst1987, Vedula2001, Garcia2006, Gonzalez01012012, Attili2019}.
A recurring observation has been that
the scalar gradient preferentially aligns with the most compressive
eigenvector of strain,
suggesting that their amplification is primarily
driven by compressive action of the local flow. 
This picture is also consistent with observed spatial
organization of scalar gradients (and dissipation) 
into sheet-like structures \cite{Vedula2001, Schumacher2005}.
Yet numerous key aspects remain unexplored. 
Prior studies have predominantly relied on unconditional
statistics, which characterize the mean-field
behavior rather than the extreme events that dominate intermittency. 
Since the dynamics of extreme events may differ 
fundamentally from globally averaged behavior, 
statistics conditioned on the local scalar dissipation 
provide a more natural framework to understand the underlying 
amplification mechanisms. 
Such conditional statistics are also valuable
in various turbulence modeling frameworks \cite{pope1994, Klimenko1999, Pitsch2000}.
Similar conditional analyses have recently provided important
insights into velocity-gradient intermittency, where conditioning on intense
vorticity or strain has revealed the dominant production mechanisms,
characteristic geometric organization, and their Reynolds-number
dependence \cite{Tsi2009, BBP2020, BPB2022}. However, comparable 
investigations for passive scalars remain severely limited.

Prior studies have also been largely restricted
to low Reynolds numbers, and the more accessible
case of $Sc = \mathcal{O}(1)$, where the viscous and
diffusive scales are comparable, as set by the 
Kolmogorov length scale $\eta_K$. The problem becomes
substantially richer at high $Sc$, where reduced 
scalar diffusivity enables scalar-gradient amplification 
down to the Batchelor scale $\eta_B = \eta_K Sc^{-1/2}$ \cite{Batchelor1959},
leading to generation of increasingly intense gradients
with $Sc$. Yet this regime has remained 
unexplored 
owing to the stringent resolution constraints
required to resolve $\eta_B$ at higher Reynolds numbers. 
An additional source of complexity arises from the imposed 
mean scalar gradient, which continuously injects scalar-variance 
and introduces anisotropic forcing into the system. 
As a result, the scalar field is known to violate local isotropy 
even at smallest scales \cite{Yeung2002, BCSY2021a, Tang2023}. 
However, its role in shaping scalar gradients 
remains poorly understood. In particular, it is unclear 
whether the most intense events are governed purely by 
local small-scale dynamics or retain memory of the 
large-scale directional forcing. 

In this work, we address these gaps through a systematic 
investigation of scalar gradient dynamics and the mechanisms 
governing the formation of extreme scalar dissipation events.
We utilize a massive direct numerical simulation (DNS) database of 
stationary isotropic 
turbulence with Taylor-scale Reynolds number $\re$ spanning
$140-1000$ on grid sizes of up to $8192^3$, 
with $Sc$ ranging from $1$ to $512$. Particular attention
is given to resolve the smallest scales and extreme events
accurately \cite{DY2010, BPBY2019}. 
We compute both unconditional statistics and 
statistics conditioned on $\chi$ to 
analyze the amplification dynamics. 
In agreement with previous studies,
the  unconditional statistics show 
that scalar gradients preferentially
align with the the most compressive eigendirection
of strain, while being nearly orthogonal to vorticity.
The alignments are greatly enhanced in regions of
intense scalar dissipation, where amplification is
carried entirely by the most compressive
eigendirection. However, the effective strain
responsible for this amplification remains of the order
of the mean-field, implying intense scalar dissipation
primarily arises from optimal alignments, rather than 
intense strain itself. 
On the other hand, the contribution from imposed mean-gradient 
is mostly negligible to the overall budget; nevertheless, 
its interaction with strain allows for direct transfer of
large-scale anisotropy to the smallest scales. 
Overall, the results support a simple physical picture
in which intense scalar dissipation is organized into 
curved sheet-like structures, embedded in shear layers between
vortex tubes, amplified by compressive strain and ultimately
arrested by molecular diffusion.

The rest of the paper is organized as follows. In 
\S~\ref{sec:setup}, we present the governing equations and DNS database. 
Relevant transport equations and background is briefly
discussed in \S~\ref{sec:bg}. 
Unconditional statistics from DNS are shown in \S~\ref{sec:uncond}
to understand the mean-field behavior, and thereafter, various statistics
conditioned on scalar dissipation rate are explored  
in \S~\ref{sec:cond}. Finally, we 
summarize our results in \S~\ref{sec:conc}.

\section{Numerical Approach and Database}
\label{sec:setup}

The data utilized in this work are 
obtained from direct numerical simulations 
(DNS) of governing conservation
equations. 
For the velocity field $\uu$, we solve
the incompressible Navier-Stokes equations: 
\begin{align}
  \frac{\partial u_i}{\partial t} + u_j\frac{\partial u_i}
  {\partial x_j} &= -\frac{\partial P}{\partial x_i} + 
  \nu\nabla^2 u_i + f_i \ , \\ 
  \frac{\partial u_i}{\partial x_i} &= 0 \ ,
  \label{eq:NS}
\end{align}
where $P$ is the kinematic pressure, $\nu$ is the kinematic 
viscosity, and $\ff$ is the large-scale forcing term
to achieve statistical stationarity \cite{EP88}.
For the passive scalar field $\theta$, we
solve the advection-diffusion equation along
with a uniform mean-gradient:
\begin{align}
  \frac{\partial \theta}{\partial t} + u_j\frac{\partial 
  \theta}{\partial x_j} = -G_j u_j + D \nabla^2 \theta,
  \label{eq:theta}
\end{align}
where $D$ is the 
scalar diffusivity. The uniform mean-gradient is taken along the
$x$-direction: $\mathbf{G} = (G,0,0)$ and provides
the forcing needed to achieve a statistically stationary 
state for the scalar \cite{pumir94, Overholt1996DirectNS}.

The DNS corresponds to the canonical setup of
isotropic turbulence in a $(2\pi)^3$ periodic domain.
The velocity field is always solved using the highly
accurate Fourier pseudospectral algorithm,
with aliasing errors controlled using a combination
of grid shifting and truncation \cite{PattOrs71, Rogallo},
giving a maximum resolved 
wavenumber $k_{\max} = \sqrt{2}N/3$, where $N$ is the 
number of grid points in each direction.

\begin{table}[htbp]
\centering
\setlength{\tabcolsep}{16pt}
\begin{tabular}{ccccccc}
\toprule
$\re$ & $Sc$ & $N_u^3$ &  $k_{\max}\eta$ & $N_\theta^3$ & $k_{\max}\eta_B$ 
& $T_{\mathrm{sim}}/T_E$ \\
\midrule
\multirow{5}{*}{140}
 & 1    & $1024^3$  & 6 & $1024^3$ & 6 & 10 \\
 & 8 & $1024^3$  & 6 & $1024^3$ & 2 & 10 \\
 & 32   & $1024^3$  & 6 & $2048^3$ & 2 & 11 \\
 & 128  & $1024^3$  & 6 & $4096^3$ & 2 & 12 \\
 & 512 & $1024^3$  & 6 & $8192^3$ & 2 & 9 \\
\midrule
\multirow{2}{*}{240}
 & 1  & $2048^3$  & 6 & $2048^3$ & 6 & 8 \\
 & 8 & $2048^3$  & 6 & $2048^3$ & 2 & 8 \\
\midrule
\multirow{2}{*}{390}
 & 1 & $4096^3$  & 6 & $4096^3$ & 6 & 3 \\
 & 8  & $2048^3$  & 6 & $8192^3$ & 4 & 6 \\
\midrule
\multirow{2}{*}{650}
 & 1 & $6144^3$  & 4.5 & $6144^3$ &  4.5 & 3 \\
 & 8 & $8192^3$  & 6   & $8192^3$ &   2  & 2 \\
\midrule
1000 & 1  & $8192^3$ & 3 & $8192^3$ & 3 & 2 \\
\bottomrule
\end{tabular}
\caption{Simulation parameters for the DNS runs utilized in the current 
work: the Taylor-scale Reynolds number $\re$, the Schmidt number $Sc$, 
the number of grid points for the velocity and scalar fields $N_v^3$ 
and $N_\theta^3$, the spatial resolution for the velocity and scalar 
fields $k_{\max}\eta$ and $k_{\max}\eta_B$, and the simulation length 
$T_{\mathrm{sim}}/T_E$ in statistically stationary state in terms of 
the large-eddy turnover time $T_E$. Cases with $N_\theta = N_u$ are 
solved using the conventional pseudospectral approach 
for both velocity and scalar, while cases with 
$N_\theta > N_u$ employ a hybrid spectral-compact difference approach 
\cite{gotoh12a, clay.cpc1, clay.cpc2}.
}
\label{tab:dns}
\end{table}

For the scalars, two different approaches are utilized
depending on $Sc$. 
For $Sc = 1$ and $8$, the conventional pseudospectral 
method is employed for both the velocity and scalar 
fields simultaneously. 
For higher $Sc$, a hybrid approach is utilized, whereby the 
velocity field is solved pseudospectrally on a grid 
resolving the Kolmogorov scale, while the scalar field 
is solved on a finer grid using compact finite 
differences to adequately resolve the Batchelor 
scale~\cite{gotoh12a, clay.cpc1, clay.cpc2}. The full database and 
simulation parameters are outlined in Table~\ref{tab:dns}.

The database spans Taylor-scale Reynolds numbers 
$\re$, which scales as $\re \sim Re^{1/2}$, in the range 
$140-1000$, on grid points of up to $8192^3$. The full Schmidt number range 
$Sc = 1$--$512$ is explored at $\re = 140$, while 
$Sc = 1$ and $8$ are considered at higher $\re$, 
allowing a systematic characterization of the effects 
of both $\re$ and $Sc$ on the scalar gradient 
dynamics. The DNS database is similar to 
that used in recent works \cite{BCSY2021a, BCSY2021b, BS2022},
and additionally, we have performed new runs
at $\re=650$, $Sc=1-8$, and $\re=1000$, $Sc=1$, expanding the 
parameter range to higher Reynolds and Schmidt numbers. 
Special attention is given to resolving the small scales
accurately \cite{buaria_jfmp}.
The resolution for velocity field is as high 
as  $k_{\max}\eta_K \approx 6$ (except at largest $\re$, where we
have $k_{\max}\eta_K \approx 3$),
whereas the resolution for the scalar 
field is $k_{\max}\eta_B \geq 2$. 
As established in prior works \cite{DY2010, BCSY2021a}, 
this resolution is sufficient
as long as the velocity field is well-resolved, which is
indeed the case for our simulations. 
All the simulations are run for several eddy turnover
times to also ensure convergence with respect to
statistical sampling.

\section{Scalar Gradient Dynamics}
\label{sec:bg}

To quantify the intensity of scalar gradient, 
we will utilize its variance $|\mathbf{g}|^2 = g_i g_i$, which is 
essentially the scalar dissipation divided by the scalar diffusivity.
Indeed, normalizing both quantities by their respective means 
shows that they are equivalent:
$ |\mathbf{g}|^2 / \langle |\mathbf{g}|^2 \rangle = \chi /\langle \chi \rangle$.
Thus, we will use scalar gradient variance interchangeably with 
scalar dissipation rate, with it being implied that they are
normalized by their mean values. 
We can obtain the transport equation for it, by taking
the dot product of Eq.~\eqref{eq:gi} with $g_i$, leading to
\begin{align}
\frac{1}{2} \frac{D (g_i g_i)}{D t} =
- g_i g_j S_{ij} - g_i G_j S_{ij} + g_i G_j R_{ij}
+ \frac{1}{2} D \, \nabla^2 (g_i g_i)  
-D \, \frac{\partial g_i}{\partial x_j} \frac{\partial g_i}{\partial x_j} \ , 
\label{eq:budget}
\end{align}
where we have decomposed the velocity gradient tensor into
its symmetric and 
skew-symmetric components: 
\begin{align}
S_{ij} = \frac{1}{2} \, (A_{ij} + A_{ji}) \ , 
\qquad
R_{ij} = \frac{1}{2} \, (A_{ij} - A_{ji}) \ ,
\label{eq:srij}
\end{align}
the strain and rotation rate tensors, respectively.
In the above equation, the first term on the right-hand side 
captures the non-linear
amplification of the scalar gradients by the local strain.
The next two terms capture the amplification arising from the imposed
mean-gradient, whereas the last two terms respectively capture the diffusive
transport (arising from the Laplacian) and the destruction of scalar dissipation,
arising from the Hessian of the scalar field \cite{Vedula2001}.

It is worth noting that the nonlinear amplification of scalar gradients
is mediated only by the strain field, as 
$g_i g_j R_{ij} = 0$  from symmetry. 
However, both strain and rotation tensor contribute to amplification
via the imposed mean gradient. 
Noting that the rotation-rate tensor $R_{ij} = -\tfrac{1}{2}
\epsilon_{ijk}\omega_k$, where $\ww = \nabla \times \uu$ is the vorticity
field, we can show that the rotation term also takes the form  
\begin{align}
g_i G_j R_{ij} = \frac{1}{2} \,  (\bg \times \ww ) \cdot \mathbf{G},
\label{eq:rotation_identity}
\end{align}
which makes explicit the interaction of scalar gradient
and vorticity vectors in amplifying gradients through the mean-gradient.
In statistically stationary 
homogeneous turbulence, two simplifications follow 
upon averaging. First, 
the diffusion term vanishes identically by homogeneity, 
$\langle\tfrac{D}{2}\nabla^2(g^2)\rangle = 0$.
Second, as derived in Appendix~\ref{app:identity}, 
it can be shown that the two
mean-gradient terms are equal upon averaging
\begin{align}
-\langle g_i G_j S_{ij}\rangle = \langle g_i G_j R_{ij}\rangle \ . 
\label{eq:mean_grad_balance}
\end{align}
However, as it will become clear in the next section, these terms
are still significantly smaller than the nonlinear 
term, i.e.,  $\langle g_i G_j S_{ij}\rangle \ll \langle g_i g_j S_{ij}\rangle$,
implying that in the stationary state 
the following production-destruction balance holds
\begin{align}
  -\langle g_i g_j S_{ij}\rangle \approx 
  D\left \langle \frac{\partial g_i}{\partial x_j}
  \frac{\partial g_i}{\partial x_j}\right \rangle \ .
  \label{eq:approx_balance}
\end{align}

It is worth emphasizing that the transport of scalar gradients 
bears a close structural 
analogy to that of vorticity. 
In statistically stationary isotropic turbulence, 
the transport of enstrophy $\Omega = \omega_i \omega_i$ 
is described by an equation identical to Eq.~\eqref{eq:budget} without
the mean-gradient terms
\footnote{with the mean-gradient term included, an analogy
can be drawn to homogeneous shear flows}
and $g_i$ replaced by $\omega_i$.
In fact, the production-destruction balance
for enstrophy is also given as: 
$\langle\omega_i\omega_j S_{ij}\rangle = 
\nu\langle \frac{\partial \omega_i}{\partial x_j}\frac{\partial \omega_i}{\partial x_j}\rangle$, 
analogous to Eq.~\eqref{eq:approx_balance} \cite{BBP2020}. 
However, a fundamental distinction remains: 
both the vorticity $\omega_i$ 
and the strain $S_{ij}$ belong to the same velocity 
field and are non-locally coupled 
\cite{Ohkitani:95, ham_pof08, BPB2020, BLW:2024},
whereas the scalar gradient is a passive field with no
feedback on the velocity gradients. 
Consequently, the underlying mechanisms and 
observed dynamics of scalar gradients
differ fundamentally from those of vorticity.

The central quantity driving the formation of extreme scalar 
dissipation is the amplification term $-g_ig_jS_{ij}$. 
Evidently, its sign and magnitude depend  
both on the orientation of $\mathbf{g}$ relative to the principal 
axes of $S_{ij}$ and the magnitudes of the corresponding 
eigenvalues. Thus, this term is most conveniently
analyzed in the eigenframe of the strain tensor $S_{ij}$ 
\cite{Ashurst1987, Tsi2009, BBP2020},
defined by the eigenvalues $\lambda_i$
and the eigenvectors $\mathbf{e}_i$, for $i = 1, 2, 3$.
By definition, $\lambda_1 \geq \lambda_2 \geq \lambda_3$
and incompressibility additionally imposes 
$\lambda_1 + \lambda_2 + \lambda_3 = 0$, implying that 
$\lambda_1 > 0$ (extensive) and $\lambda_3 < 0$ (compressive). 
The intermediate eigenvalue $\lambda_2$ is well-known to be 
positive on average, consistent with the energy cascade from large to small
scales, and amplification of gradients \cite{Betchov56, Ashurst1987}. 
In the eigenframe of strain tensor, the scalar gradient amplification
term can be written as 
\begin{align}
 g_ig_jS_{ij}  = 
  |\bg|^2 \, \lambda_i \, (\mathbf{e}_i \cdot \mathbf{\ghat})^2 \ , 
  \label{eq:Ps_decomp}
\end{align}
where $\ghat = \mathbf{g} /|\mathbf{g}|$ is the unit vector. 
We proceed to analyze various statistics pertaining to scalar gradient
amplification in the following section.

\section{Gradient correlations: Unconditional statistics}
\label{sec:uncond}

We begin by examining unconditional correlations between scalar
and velocity gradients which provide a baseline description
of amplification dynamics. 
To allow meaningful comparisons across different Reynolds and Schmidt numbers,
the velocity gradient part of statistics
is non-dimensionalized by the Kolmogorov time 
scale $\tau_K$
\begin{align}
    \tau_K = \left(\nu/\langle\epsilon\rangle\right)^{1/2} \ ,
    \qquad {\rm with} \quad \epsilon= 2\nu S_{ij}S_{ij}  \ , 
\end{align}
being the energy dissipation rate,
whereas scalar gradient part is non-dimensionalized by
its mean amplitude, as quantified the $L^2$-norm: 
$\langle g_i g_i \rangle^{1/2}$. For convenience,
we also define here the enstrophy, the square-norm
of vorticity
\begin{align}
 \Omega = \omega_i \omega_i \ , 
\end{align}
though it will be first used in \S~\ref{sec:cond}.

\subsection{Alignments}
\label{subsec: alignments_uncond}

\begin{figure}[ht]
\begin{center}
\includegraphics[width=0.5\textwidth]{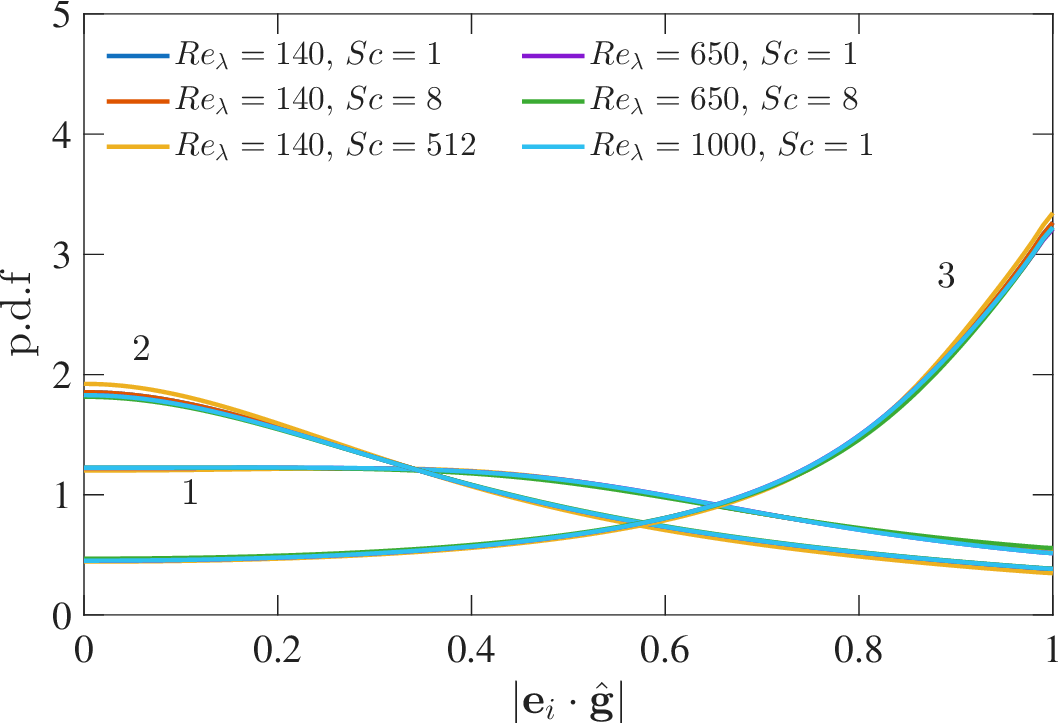}
\end{center}
\caption{Probability density function of the alignment cosines
between the scalar gradient unit vector $\ghat$ and strain eigenvectors 
$\mathbf{e}_i$, for $i = 1,2,3$, at different $\re$ and $Sc$  
listed in Table~\ref{tab:dns}. Not all cases are shown, as 
the distributions collapse and show no dependence on 
either $\re$ or $Sc$.}
\label{fig:pdf_eig}
\end{figure}

To analyze the alignment of the scalar gradient 
with the strain eigenvectors,  Fig.~\ref{fig:pdf_eig}
shows the probability density functions (PDFs) 
of the alignment cosines $|\mathbf{e}_i\cdot\ghat|$, for all
$\re$ and $Sc$ shown in 
Table~\ref{tab:dns}. Consistent with earlier
observations at lower $\re$ and $Sc\sim 1$ \cite{Ashurst1987, Vedula2001}, 
we find that scalar gradient
preferentially aligns with the most compressive eigenvector $\mathbf{e}_3$,
(as indicated by the peak of PDF near unity),
and has the tendency to be orthogonal
to the intermediate eigenvector $\mathbf{e}_2$. 
The alignment with $\mathbf{e}_1$ very weakly peaks
at $0$, but essentially appears close to uniform. 
Remarkably, the three alignment distributions show essentially 
no dependence on either $\re$ or $Sc$, indicating that the orientation 
of scalar gradients within the local strain eigenframe is 
a robust kinematic feature of scalar mixing. 
In fact, similar alignment characteristics have also been 
observed in turbulent premixed and nonpremixed flames 
\cite{Hamlington2011, Attili2019},
where the alignments converges to that of incompressible turbulence 
at high Reynolds numbers.

A convenient quantitative measure of these alignments is provided by 
the second moment $\langle(\mathbf{e}_i\cdot\ghat)^2 \rangle$. This quantity
ranges from $0$, corresponding to perfect orthogonality, to $1$, corresponding
to perfect alignment, with $1/3$ for a uniform distribution of the PDF, corresponding
to no preferential alignment. 
In addition, it has the nice property that contributions in all three
eigendirections also add up to unity, i.e., 
$\sum_{i=1}^3 \langle(\mathbf{e}_i\cdot\ghat)^2 \rangle = 1$.
We have listed the second moments for all runs 
in Table~\ref{tab:stats_basics}. For each case, they 
are approximately in the ratio $0.26:0.20:0.53$
for $i = 1,2,3$, reinforcing the observations in 
Fig.~\ref{fig:pdf_eig}.

\begin{figure}
\centering
\includegraphics[height=0.32\textwidth]{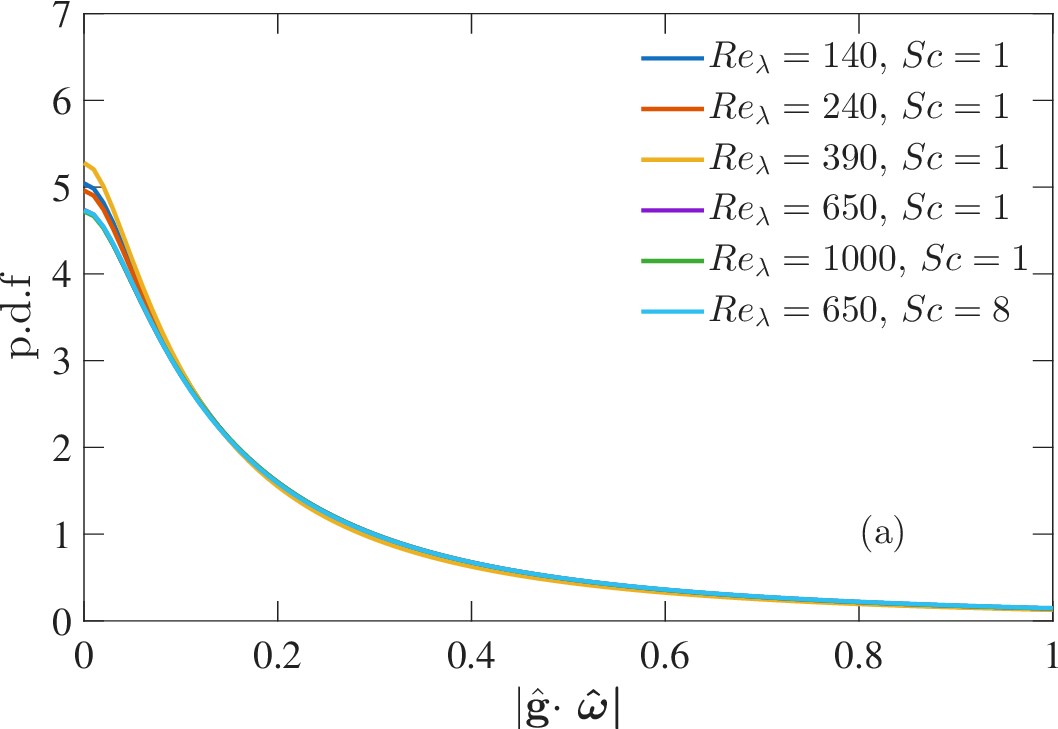} \ \ \ \ \ \ \  
\includegraphics[height=0.32\textwidth]{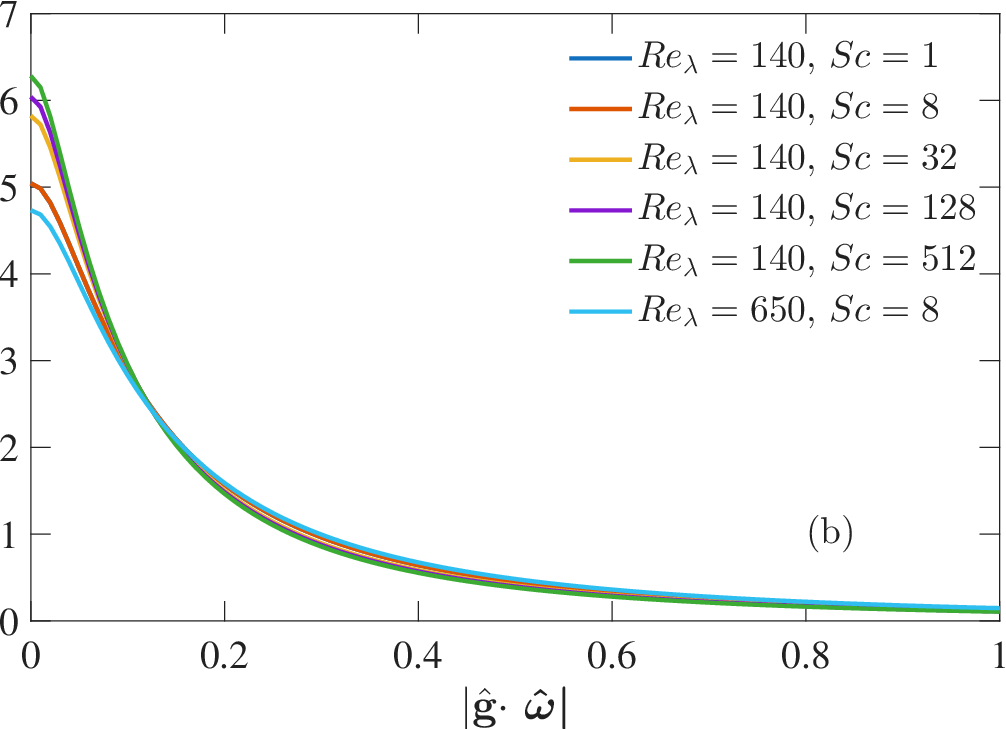}
\caption{Probability density functions of the alignment cosine 
between the scalar gradient vector $\ghat$ and vorticity unit vector 
$\omhat$ for (a) various $\re$ at $Sc=1$ and (b) various $Sc$ at $\re =140$, 
with the $\re = 650$, $Sc = 8$ case common to both.}
\label{fig:pdf_sgw}
\end{figure}

Figure~\ref{fig:pdf_sgw}  shows the PDFs of 
$|\ghat\cdot\omhat|$, the alignment cosine between 
the scalar gradient and vorticity vectors, for 
varying $\re$ (panel a) and varying $Sc$ (panel b).
For all cases, the distributions are strongly peaked near zero  
showing that scalar gradient is preferentially orthogonal
to vorticity throughout the flow. This tendency is quantified
by the second moment  $\langle(\ghat\cdot
\omhat)^2\rangle$ listed in Table~\ref{tab:stats_basics}.
The values remain small (approximately 0.1) for all cases, 
with negligible $\re$ dependence.
However, at $\re=140$, there is a weak but systematic dependence on $Sc$,
with the second moment slowly decreasing as $Sc$ increases,
consistent with the behavior of PDFs in Fig.~\ref{fig:pdf_sgw}b.
Since no comparable $Sc$-dependence is observed at higher $\re$, 
this variation appears to be a low-$\re$ effect.
At sufficiently high $\re$, the alignments appear to 
approach a robust state in which scalar gradient and vorticity vector 
remain nearly orthogonal, with no sensitivity to further changes in $\re$ or $Sc$.

This observed near-orthogonality between the scalar gradient 
and vorticity is perhaps to be expected. 
It is well known that vorticity 
preferentially aligns with $\mathbf{e}_2$, the intermediate eigenvector of strain
\cite{Ashurst1987, Tsi2009, BBP2020}. 
Given that scalar gradient appears to preferentially align
with $\mathbf{e}_3$, which is by construction perpendicular to 
$\mathbf{e}_2$, the orthogonality of scalar gradient and
vorticity follows kinematically from their respective
orientations within the eigenframe of strain.

\subsection{Scalar gradient amplification statistics}
\label{subsec:scg_stats}

Following the alignment results,  
Table~\ref{tab:stats_basics} next reports unconditional averages 
pertaining to the amplification terms. 
It is worth noting, that in addition to the alignments,
the magnitude of strain and its eigenvalues also play 
a central role in determining the net amplification 
rate of scalar gradients. However, the unconditional
results for strain eigenvalues have been previously
reported in \cite{BBP2020}, and we do not repeat them here
again. Though it is worth recounting
that the non-dimensional expectations $\langle \lambda_i^2 \rangle \tau_K^2$,
are approximately in the ratio $0.18: 0.02: 0.29$ for $i=1,2,3$,
virtually independent of $\re$. Note, by definition,   
$\langle \lambda_i \lambda_i \rangle \tau_K^2 = \tfrac{1}{2}$.
This result essentially establishes that the intermediate eigenvalue 
is a very small magnitude, whereas the third (compressive) eigenvalue dominates.
Since scalar gradient  also preferentially aligns with the third eigenvector,
the expectation would be that scalar gradient amplification is predominantly
driven by the third eigendirection of strain tensor.

\begin{table}[htbp]
\centering
\setlength{\tabcolsep}{8pt}
\begin{tabular}{ccccccc}
\toprule
$\re$ & $Sc$ & $\avg{(\mathbf{e}_i\cdot\ghat)^2}$ & $\avg{(\ghat\cdot\hat{\boldsymbol{\omega}})^2}$ & $-\dfrac{\avg{\lambda_i(\mathbf{e_i}\cdot \bg)^2}\tau_K}{\avg{g_i g_i}}$
  & $-\dfrac{\avg{g_ig_jS_{ij}}\tau_K}{\avg{g_i g_i}}$ & $\dfrac{\langle g_i G_j S_{ij} \rangle}{\langle g_i g_j S_{ij} \rangle}$ \\
\midrule
\multirow{5}{*}{140}
& 1 & $0.266 : 0.205 : 0.529$ & 0.107 & $-0.084:-0.01:0.428$ & 0.334 & 5.25e-03 \\
& 8 & $0.263 : 0.201 : 0.536$ & 0.097 & $-0.084:-0.01:0.417$ & 0.323 & 8.58e-04 \\
& 32 & $0.261 : 0.197 : 0.542$ & 0.087 & $-0.086:-0.01:0.417$ & 0.321 & 1.82e-04 \\
& 128 & $0.262 : 0.195 : 0.543$ & 0.085 & $-0.088:-0.01:0.415$ & 0.317 & 4.58e-05 \\
& 512 & $0.263 : 0.194 : 0.543$ & 0.081 & $-0.089:-0.01:0.417$ & 0.318 & 1.47e-05 \\
\midrule
\multirow{2}{*}{240}
& 1 & $0.262 : 0.205 : 0.533$ & 0.105 & $-0.087:-0.01:0.444$ & 0.347 & 2.05e-03 \\
& 8 & $0.262 : 0.203 : 0.535$ & 0.099 & $-0.091:-0.01:0.432$ & 0.331 & 3.38e-04 \\
\midrule
\multirow{2}{*}{390}
& 1 & $0.261 : 0.204 : 0.535$ & 0.103 & $-0.086:-0.01:0.439$ & 0.343 & 6.72e-04 \\
& 8 & $0.259 : 0.199 : 0.542$ & 0.09 & $-0.089:-0.01:0.438$  & 0.339 & 1.49e-04 \\
\midrule
\multirow{2}{*}{650}
& 1 & $0.260 : 0.203 : 0.535$ & 0.103 & $-0.093:-0.01:0.472$  & 0.369 & 3.67e-04 \\
& 8 & $0.264 : 0.205 : 0.531$ & 0.103 & $-0.087:-0.01:0.425$  & 0.328 & 7.29e-05 \\
\midrule
\multirow{1}{*}{1000}
& 1 & $0.260 : 0.203 : 0.535$ & 0.103 & $-0.091:-0.01:0.458$ & 0.357 & 1.11e-04 \\
\bottomrule
\end{tabular}
\caption{Unconditional statistics pertaining to scalar-gradient amplification 
based on Eq.~\eqref{eq:budget},
for all the $\re$ and $Sc$ cases listed in Table~\ref{tab:dns}. 
The table reports the alignment of the scalar gradient with 
the strain eigenvectors and vorticity unit vector, the decomposition of 
the amplification term $-\avg{g_ig_jS_{ij}}\tau_K/\avg{g_ig_i}$ 
into the three strain eigendirections $i = 1, 2, 3$, and the 
relative magnitude of the mean-gradient production 
$\avg{g_iG_jS_{ij}}$ compared to the nonlinear amplification 
$\avg{g_ig_jS_{ij}}$.  
}
\label{tab:stats_basics}
\end{table}

We first consider the nonlinear amplification term 
$-\langle g_i g_j S_{ij}\rangle$, 
non-dimensionalized by $\tau_K / \langle g_i g_i \rangle$,
listed in Table~\ref{tab:stats_basics}.
This quantity is approximately constant
for all cases, albeit with a weak $Sc$ dependence at the lowest $\re=140$. 
At higher $\re$ and $Sc$, it appears to approach a 
universal value close to
$0.33$, up to minor statistical variations.
The decomposition of this term into contributions from three
strain eigendirections is also reported in Table~\ref{tab:stats_basics}.
The three contributions appear to asymptote
to  $-0.09 : -0.01 : 0.43$ for $i=1,2,3$, respectively. 
Thus, the net amplification of scalar gradients
is predominantly driven by the most compressive eigendirection of strain.
The first eigendirection contributes
to some mild stretching (attenuation), whereas the 
intermediate direction has a negligible contribution.
The above results essentially confirm prior expectations
and point to geometric organization consistent with 
sheet-like scalar gradient structures \cite{Schumacher2006}. 
We will revisit this picture in greater detail 
in the next section
using conditional statistics, which isolate the dynamics of
increasingly intense scalar-gradient events, rather than 
averaging over the full flow.

It is worth noting that the mean amplification
term $-\langle g_i g_j S_{ij}\rangle$ can be more generally interpreted
through the mixed velocity-scalar gradient tensor  
$\langle g_i g_j A_{kl}\rangle$. Since the tensor is entirely composed
of gradients, one can show all its components can be prescribed by a single
component under the assumption of local isotropy 
\begin{align}
\langle g_i g_j A_{kl} \rangle
= \frac{3}{4}\,\langle g_1^2 A_{11}\rangle
\left(  - \frac{2}{3} \delta_{ij} \delta_{kl}
+ \delta_{ik} \delta_{jl}  + \delta_{il} \delta_{jk}  \right).
\label{eq:Mijkl_final}
\end{align}
A derivation for this is provided in 
Appendix~\ref{app:isotropic_tensor}. 
Since $g_i g_j S_{ij} = g_i g_j A_{ij}$ from symmetry, 
the use of the above isotropic form gives: 
$\langle g_i g_j S_{ij}\rangle = \langle g_i g_j A_{ij}\rangle 
= \frac{15}{2}\langle g_1^2 A_{11} \rangle$, 
which upon non-dimensionalization leads to the result-
\begin{align}
    \frac{\langle g_i g_j S_{ij}\rangle\tau_K}{\langle g_i g_i\rangle}
    = \frac{5}{2\sqrt{15}}\,\mathcal{S}_{u\theta} \ , \qquad
    {\rm where} \ \ \  \mathcal{S}_{u\theta} = \frac{\langle g_1^2 A_{11}\rangle}
{\langle g_1^2\rangle\langle A_{11}^2\rangle^{1/2}}\ , 
    \label{eq:amp_skewness_relation}
\end{align}
is the mixed velocity-scalar gradient skewness \cite{Yeung02, Antonia2003}. 
Here, we have also utilized the isotropic relations 
$15\langle A_{11}^2\rangle = \langle \epsilon \rangle / \nu = 1/\tau_K^2$, 
and $\langle g_1^2\rangle = \frac{1}{3}\langle g_i g_i\rangle$. 

It is well known that $\mathcal{S}_{u\theta}$ is negative
and captures the forward cascade of scalar variance from large
to small scales \cite{MY.II}, analogous to 
negative skewness of longitudinal velocity gradients and forward
cascade of kinetic energy \cite{Betchov56}.  
The mixed derivative skewness has been examined
in previous studies and is expected to approach a
constant value at sufficiently high $\re$ and $Sc$ 
\cite{Yeung2002, Antonia2003, Tang2016, Tang2025}.
Substituting $-0.33$ for the 
non-dimensional amplification term in 
Eq.~\eqref{eq:amp_skewness_relation} leads
to the result $\mathcal{S}_{u\theta} \approx -0.51$, 
which is in near-perfect agreement with previous studies
at lower $\re$ and $Sc$ \cite{Yeung2002,Tang2025}. 


\paragraph*{Role of imposed mean-gradient:}
Finally, in the last column of Table~\ref{tab:stats_basics} 
we show the quantity $\langle g_i G_j S_{ij}\rangle/\langle 
g_i g_j S_{ij}\rangle$, which captures the relative amplification
arising from the mean-gradient term versus the nonlinear term. 
It can be seen that this ratio always remains small,
becoming increasingly smaller as either $\re$ 
or $Sc$ increases. Thus, at the level of global mean budget, the scalar gradient
amplification is overwhelmingly dominated by the nonlinear
term, with the direct contribution from imposed mean gradient
being negligible.

\begin{figure}
\centering
\includegraphics[width=0.5\textwidth]{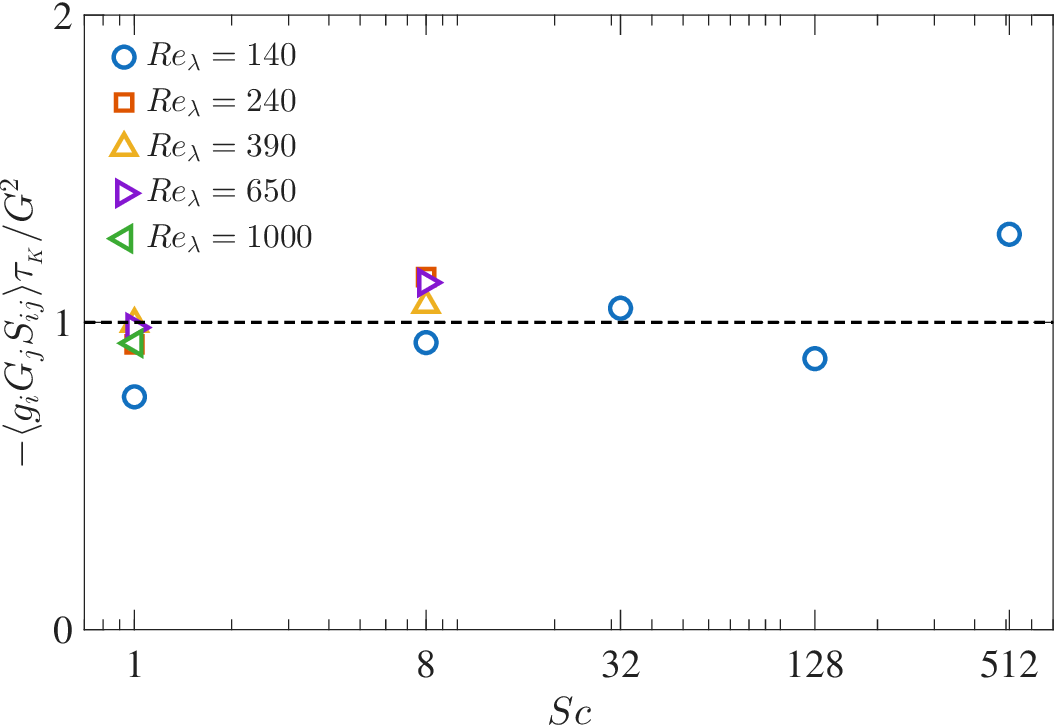}
\caption{
Non-dimensionalized mean-gradient production term
$-\langle g_iG_jS_{ij}\rangle\tau_K/G^2$ as a function of $Sc$ 
for different $\re$. The horizontal dashed line at unity corresponds to the scaling prediction in Eq.~\eqref{eq:meangrad_scale}. 
}
\label{fig:normalized_mean_gradient}
\end{figure}

Since the mean-gradient is constant and 
imposed along the Cartesian coordinate direction, 
we can easily understand this behavior by writing
$\langle g_i G_j S_{ij}\rangle = \langle g_i S_{ij}\rangle G_j$.
From previous observations on alignment, one may write
$ - \langle \mathbf{S} \cdot \bg \rangle \approx g \lambda_3 \mathbf{e}_3$.
However, the orientation of eigenvector $\mathbf{e}_3$ with respect 
to any Cartesian coordinate direction is essentially random. 
As a result, when projected along the mean-gradient, the ensuing
cancellation leads to the expectation 
\begin{align}
    \langle g_i G_j S_{ij}\rangle \sim \frac{G^2}{\tau_K} \ , 
    \label{eq:meangrad_scale}
\end{align}
where $1/\tau_K$ captures the mean strength of strain. 
We test this result in 
Fig.~\ref{fig:normalized_mean_gradient}.
Remarkably, we observe all data points are scattered 
around a value of $1$ with no systematic dependence
on either $\re$ and $Sc$, confirming 
that the mean-gradient production term is set solely 
by $G$ and $\tau_K$.

It is worth noting that the corresponding
result shown in Table~\ref{tab:stats_basics} displayed
both $\re$ and $Sc$ dependence, which is linked
to the choice of normalization. 
Since $\langle g_i g_j S_{ij} \rangle \sim \langle g_i g_i\rangle /\tau_K$ 
and $\langle g_i G_j S_{ij} \rangle \sim G^2/\tau_K$,
their ratio as given in Table~\ref{tab:stats_basics} 
is proportional to $G^2 / \langle g_i g_i \rangle$.
It is easy to see that this ratio can be related to scalar dissipation
anomaly, leading to a dependence of the form $1/(\re^2 Sc)$,
which is perfectly consistent with the data in Table~\ref{tab:stats_basics}.

\section{Intense scalar dissipation: conditional statistics} 
\label{sec:cond}

The unconditional statistics presented in the previous 
section establish the mean dynamics of scalar gradients. 
In this section, to 
isolate the dynamics responsible for the formation of extreme events, 
we will condition the statistics on the local
scalar dissipation rate, specifically $\chi /\langle \chi \rangle$. 
To capture the full range of variation in scalar dissipation,  
the conditioning is performed using logarithmically spaced bins, with eight bins 
per decade. The strategy follows previous conditional analyses
of extreme velocity gradients based on enstrophy and energy dissipation 
\cite{BBP2020,BPB2022}, allowing us to draw parallels when appropriate.

\begin{figure}[p]
\centering
\begin{tabular}{@{}c@{\hspace{0.04\textwidth}}c@{}}
\includegraphics[width=0.45\textwidth]{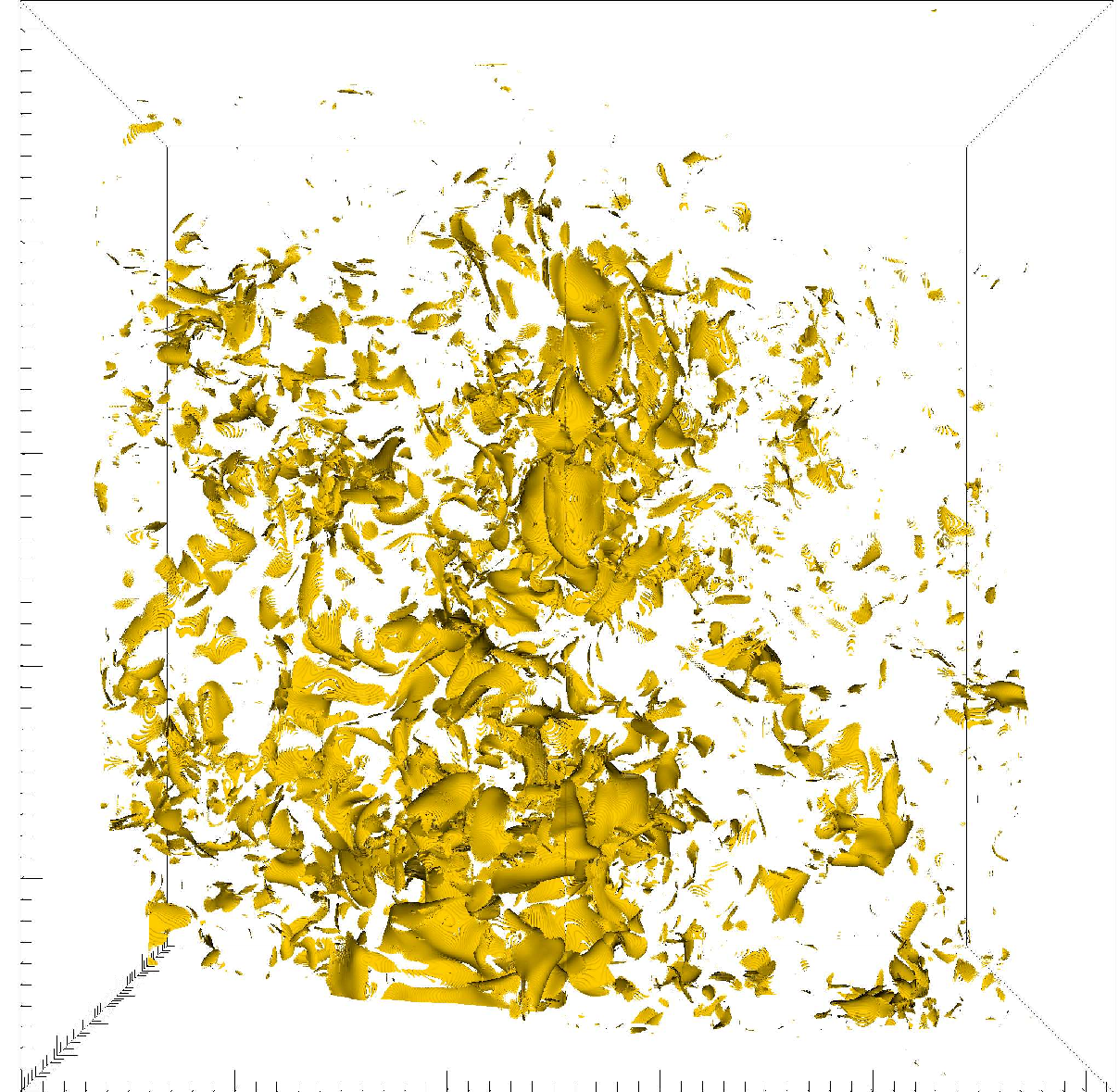} &
\raisebox{0.73cm}{\includegraphics[width=0.35\textwidth]{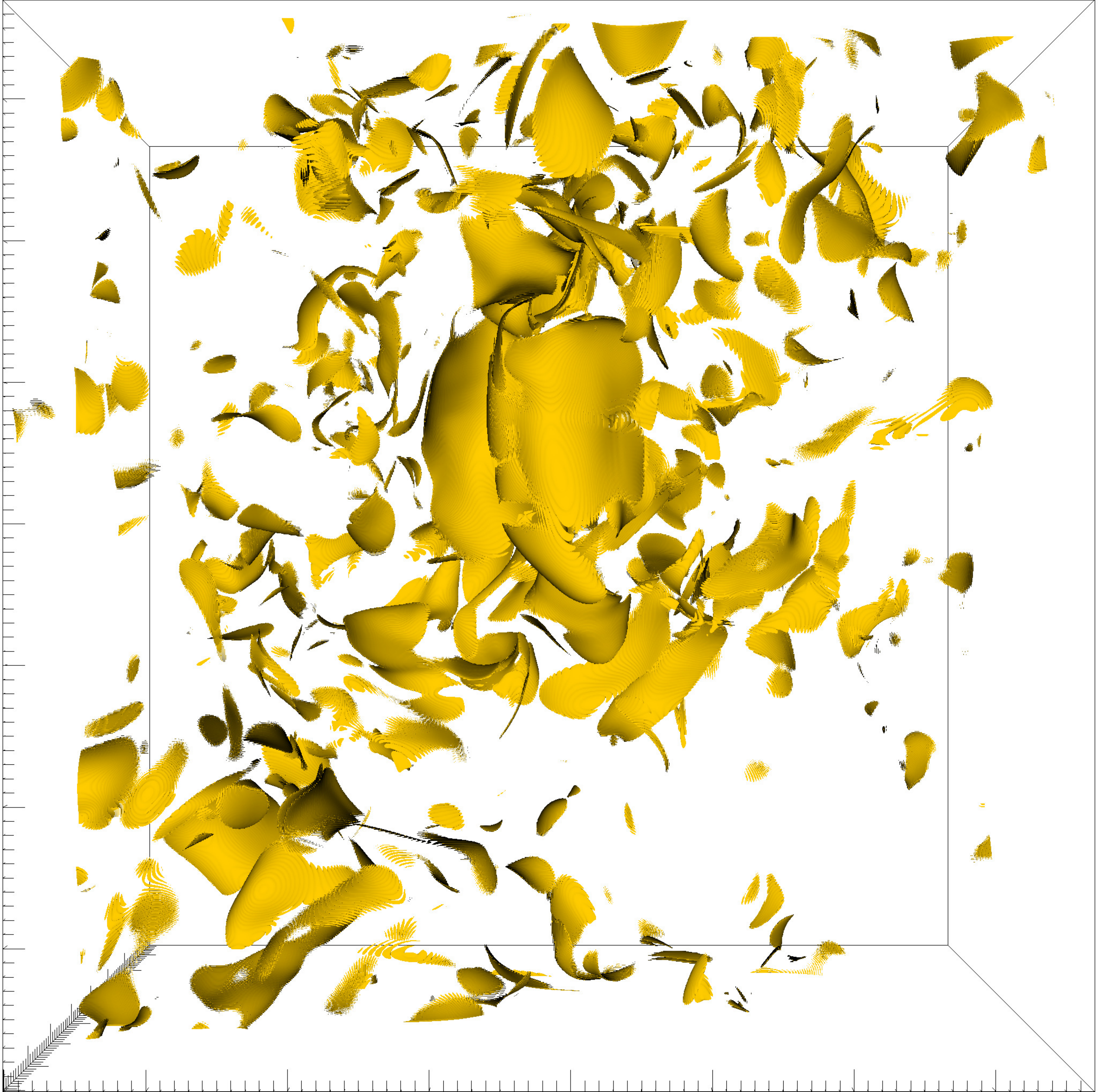}} \\
(a) $(2049)^3$, $C_\chi=75$ & \smash{\raisebox{0.65cm}{(b) $(769)^3$, $C_\chi=150$}}
\end{tabular}

\vspace{1em}

\hspace*{0.01\textwidth}
\begin{tabular}{@{}c@{\hspace{0.04\textwidth}}c@{\hspace{0.04\textwidth}}c@{}}
\includegraphics[width=0.254\textwidth]{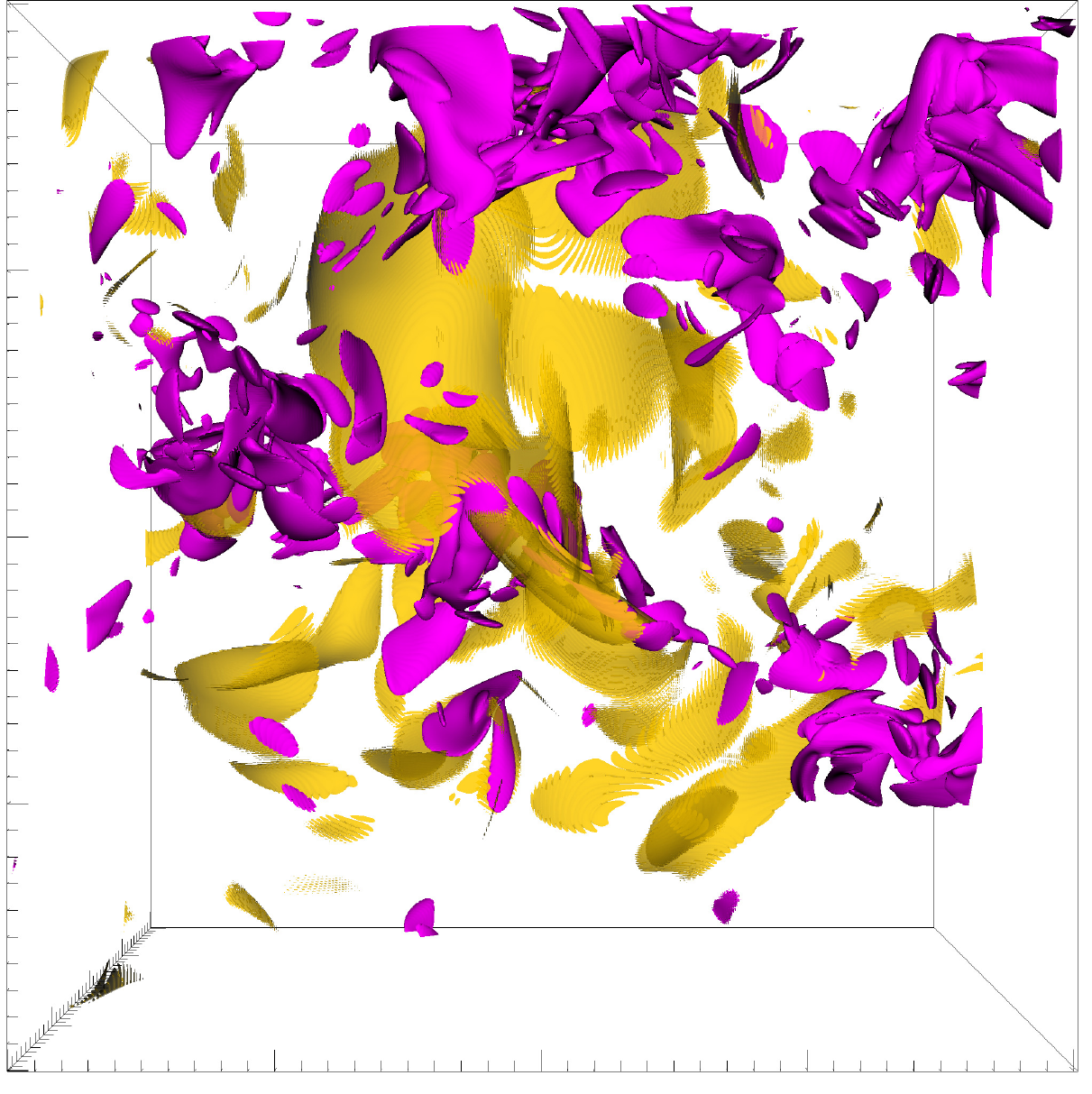} &
\includegraphics[width=0.254\textwidth]{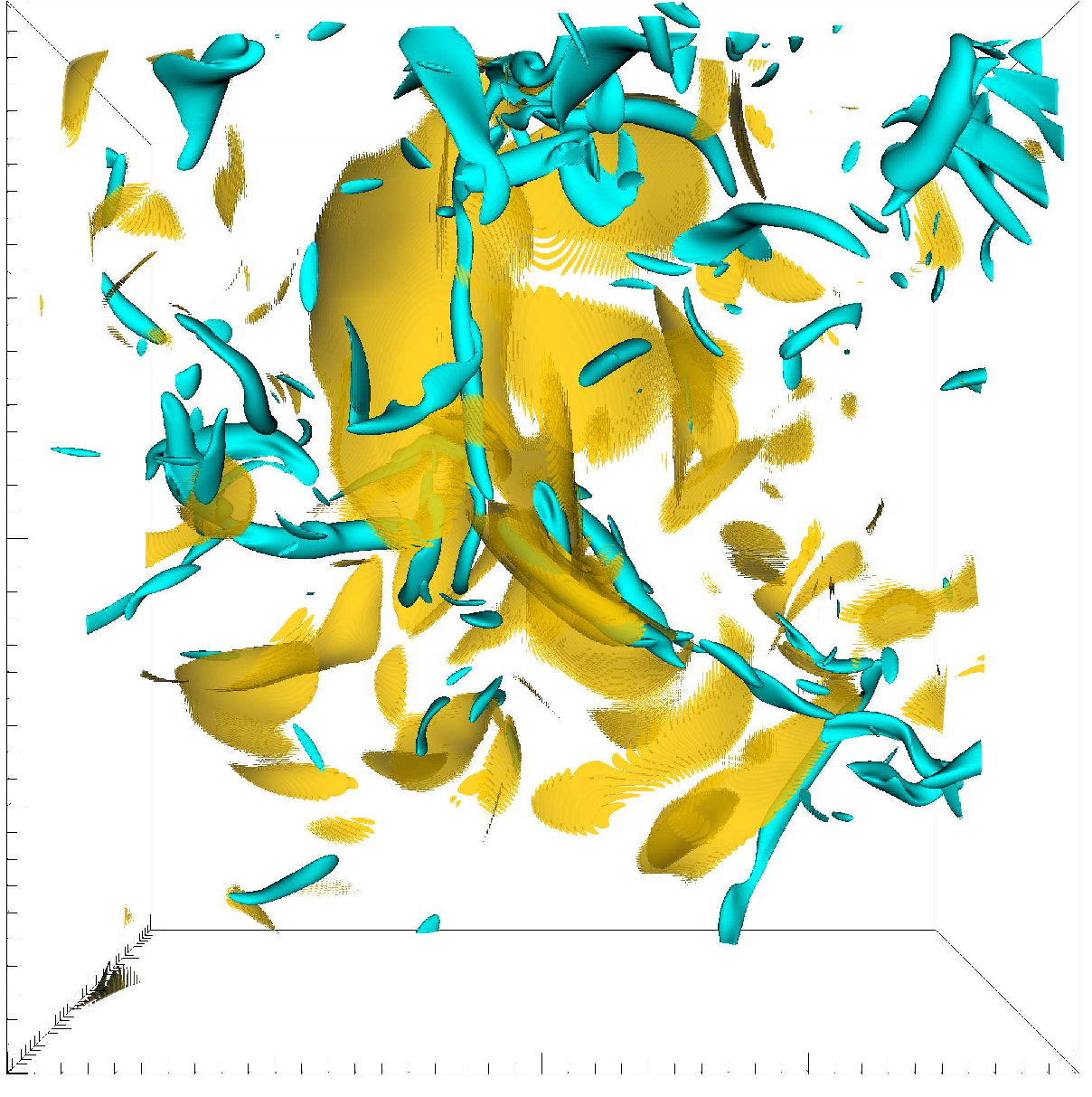} &
\includegraphics[width=0.254\textwidth]{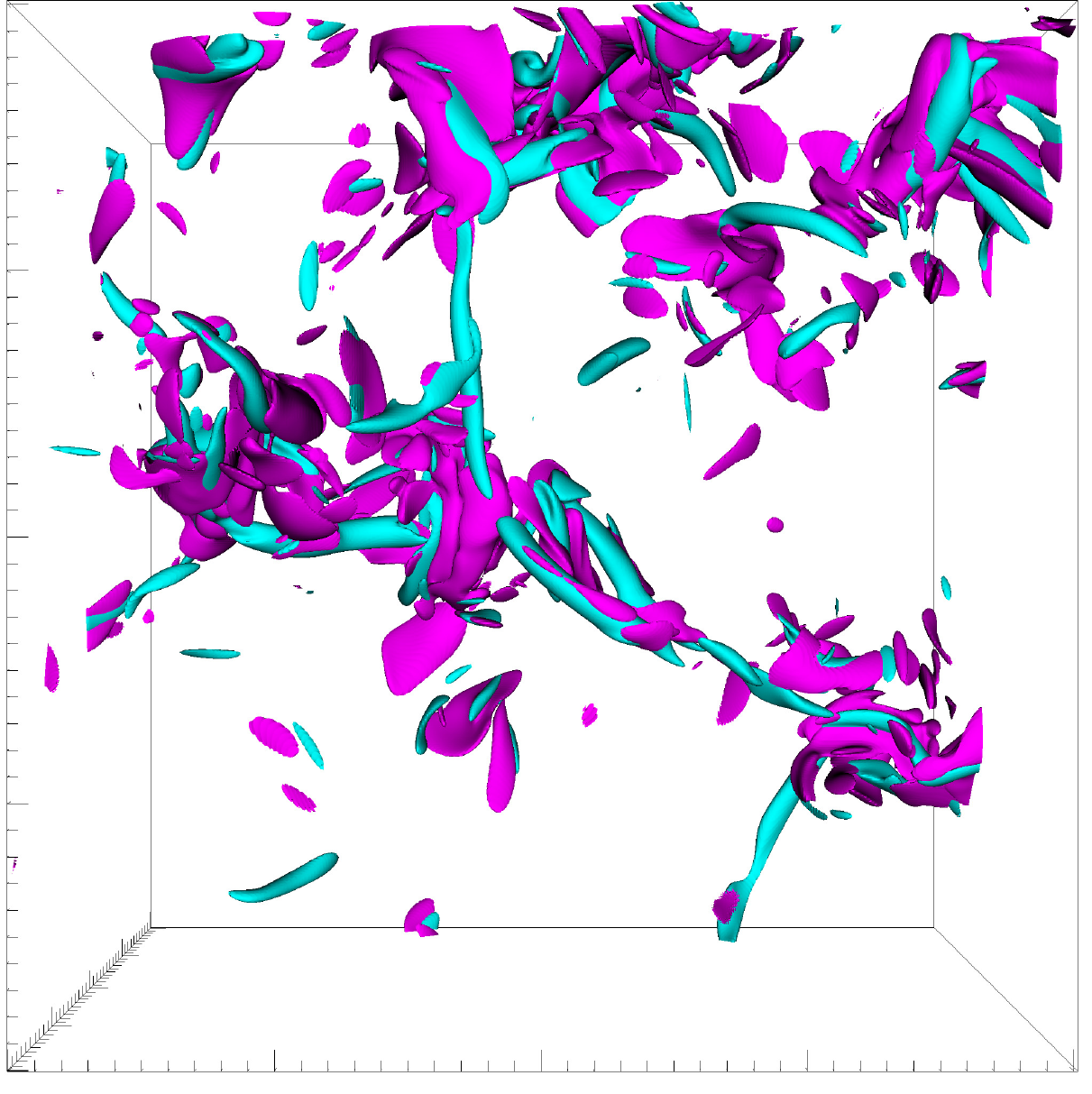} \\
(c) $(401)^3$, $C_\chi=200$, $C_\epsilon=10$ &
(d) $(401)^3$, $C_\chi=200$, $C_\Omega=20$ &
(e) $(401)^3$, $C_\epsilon=10$, $C_\Omega=20$
\end{tabular}

\vspace{1em}

\hspace{0.045\textwidth}
\includegraphics[width=0.88\textwidth]{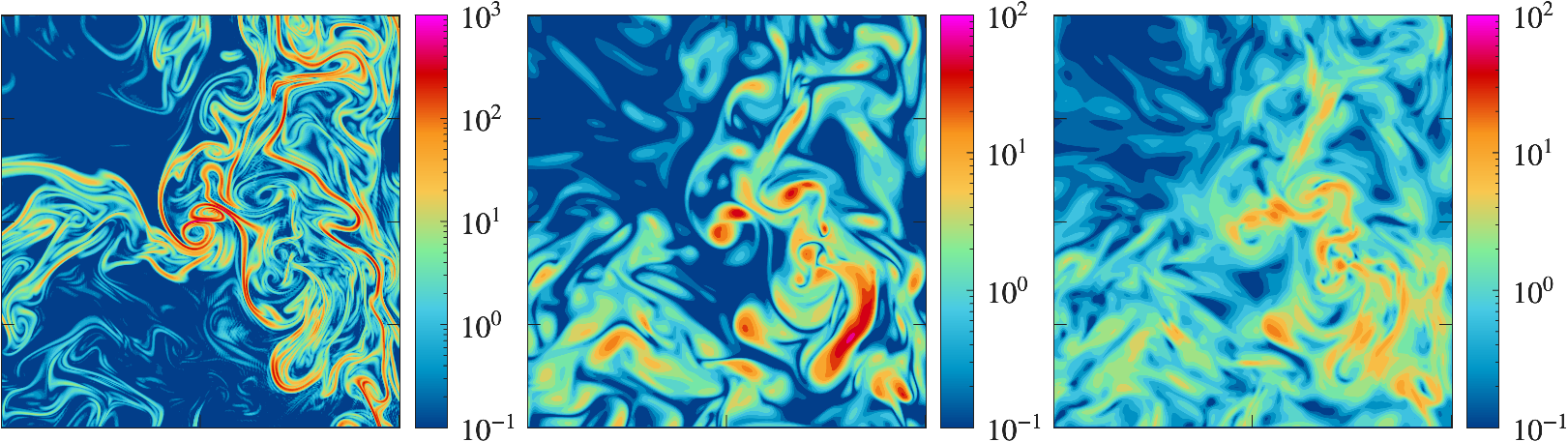}\\[0.3em]
\makebox[\textwidth]{%
\hspace{-0.01\textwidth}%
\makebox[0.38\textwidth]{(f) $(401)^2$, $\chi$}%
\hspace{-0.05\textwidth}%
\makebox[0.32\textwidth]{(g) $(401)^2$, $\Omega$}%
\hspace{-0.02\textwidth}%
\makebox[0.32\textwidth]{(h) $(401)^2$, $\varepsilon$}%
}
\caption{
Visualizations of instantaneous fields from the $8192^3$ simulation 
at $Re_\lambda=650$, $Sc=8$, centered on the point corresponding
to global maxima of scalar dissipation $\chi$ in the snapshot.
Panels (a) and (b) 
show the isosurfaces of scalar dissipation $\chi$ at progressively higher contour 
thresholds, shown in successively smaller subdomain centered on the same extreme event. 
Joint isosurfaces from a $(401)^3$ subdomain showing the spatial relationship between scalar dissipation, energy dissipation, and enstrophy are shown in panels: (c) $\chi$ (yellow) with $\epsilon$ (magenta), (d) $\chi$ (yellow) with $\Omega$ (cyan), and (e) $\epsilon$ (magenta) with $\Omega$ (cyan). Panels (f)-(h) shows the corresponding 2D slices at the midplane of
the same subdomain with contours of $\chi$, $\Omega$, and $\epsilon$, respectively. The domain size and contour thresholds $C$ used for each isosurface visualization are indicated below the corresponding panels.
The grid spacing in the domain corresponds 
to $\Delta x \approx \eta_K/2 \approx 1.4 \eta_B$, where $\eta_K$ and $\eta_B$ 
are the Kolmogorov and Batchelor length scales, respectively.
}
\label{fig:contours}
\end{figure}

\subsection{Structure of scalar and velocity gradients}
\label{subsec:structure}

We begin by studying instantaneous flow visualizations to develop some intuition
on spatial organization of extreme scalar and velocity gradients. 
Figure~\ref{fig:contours} shows the structure of scalar dissipation,
enstrophy (square-norm of vorticity), and energy dissipation from a characteristic
snapshot at $Re_\lambda = 650$, $Sc = 8$. All quantities are normalized by their
respective means. The most intense scalar dissipation event lies at
the center of each domain. 
Panels a and b show only the isosurfaces  of intense scalar dissipation,
at contour thresholds of $75$ and $150$ times the mean respectively,
illustrating the organization of scalar dissipation into convoluted sheet-like
structures -- in line with earlier observations at lower 
$\re$ \cite{Vedula2001, Schumacher2005, Watanabe2004NJP}. 
In Fig.~\ref{fig:contours}c, the isosurfaces of scalar dissipation (yellow) 
and energy dissipation (red) are shown together, suggesting that intense
scalar dissipation is accompanied by less intense strain. Note
that the domain captures the most intense scalar dissipation; however, the corresponding
maximum intensity of  energy dissipation in the same domain is more than an order
of magnitude smaller. 

Figure~\ref{fig:contours}d  shows the isosurfaces of scalar dissipation and enstrophy.
In this case, we observe the classical vortex tubes for enstrophy 
\cite{Jimenez93, Ishihara07, BPBY2019}, with sheets of scalar dissipation
often surrounding them.  Panel e shows the classical
structure of energy dissipation and enstrophy, with vortex tubes surrounded by
sheet of intense strain, providing a reference for previous cases 
\cite{Jimenez93, moisy:2004, BPBY2019}.
Thus, from these visualizations, a simple picture emerges:
in the velocity field, intense vorticity resides in tubes, with intense
sheet-like strain surrounding it, and within these shear layers
around vortex tubes, strain also acts to amplify scalar gradients.
It is worth noting that while the amplification of vorticity by strain
is nonlocal \cite{Ohkitani:95, BP2021}, 
the amplification of scalar gradients by strain
is entirely local. As a result, unlike intense vorticity and strain, 
intense scalar gradients and strain are often co-located.

The above picture is further corroborated by 2D contour fields  
shown in Fig.~\ref{fig:contours}f-h, corresponding to mid-plane 
of the domain shown in panels c-e. 
Using enstrophy in panel g -- which shows several vortex cores -- 
as a reference, it can be observed that both intense scalar dissipation
and energy dissipation in panels f and h, respectively, are confined to thin
winding shear layers surrounding the vortex cores. 
However, in these shear layers, the corresponding
intensity of scalar dissipation is much higher than that of energy dissipation
(and also enstrophy in nearby vortex cores). 
As a comparison, the most intense energy dissipation and enstrophy events
at this $\re$ can be several thousand times the mean \cite{BPBY2019, BP2022}.
Thus, it appears the most intense scalar dissipation events are not
necessarily produced by the most intense strain.

\begin{figure}
\centering
\includegraphics[height=0.32\textwidth]{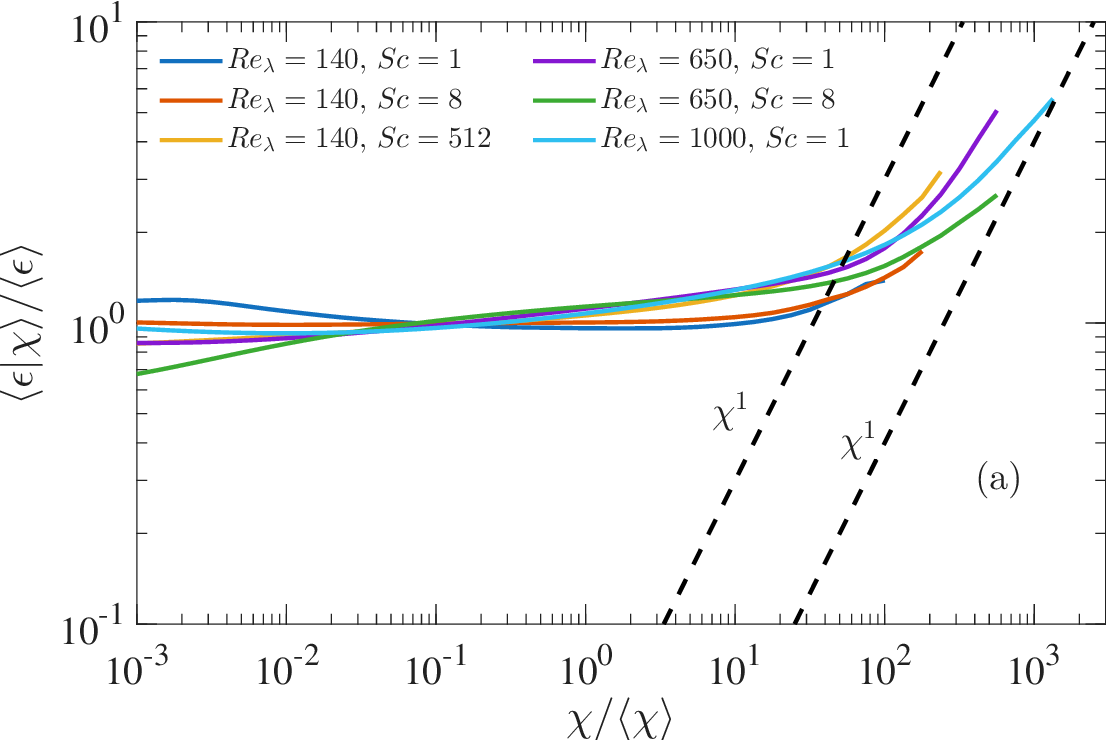} \ \ \ 
\includegraphics[height=0.32\textwidth]{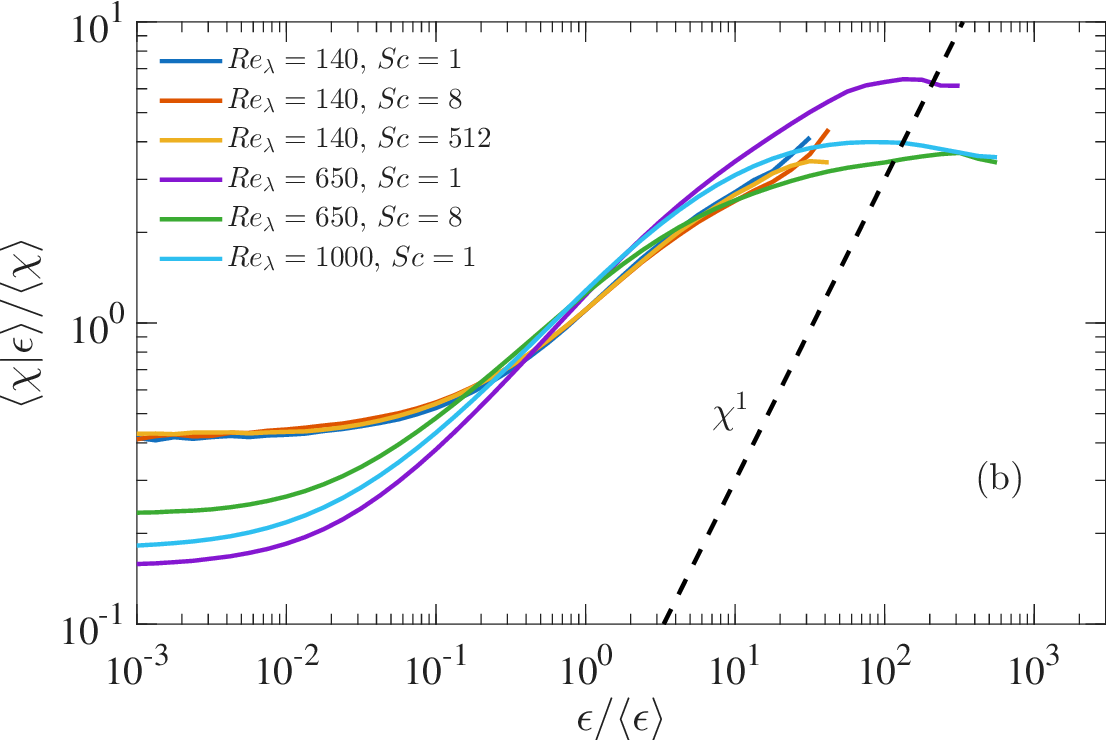}
\caption{Conditional expectations of (a) energy dissipation on scalar dissipation and 
(b) scalar dissipation on energy dissipation, normalized by their respective unconditional 
expectations, for various $\re$ and $Sc$.}
\label{fig:diss}
\end{figure}

To precisely quantify the picture drawn above,  
we first analyze in Fig.~\ref{fig:diss}, the mutual conditional
statistics of scalar and energy dissipation, with panel a
showing the conditional expectation $\langle \epsilon | \chi \rangle$
and panel b showing $\langle \chi | \epsilon \rangle$,
for various $\re$ and $Sc$. 
Focusing first on Fig.~\ref{fig:diss}a, we observe that
scalar dissipation events for $\chi /\langle \chi \rangle \lesssim 10$
are nominally governed by the mean energy dissipation,
but for more intense events the energy dissipation also grows
more intense, approaching a simple linear relation
$\langle \epsilon | \chi \rangle \sim \chi$ for the most intense events.
However, note that even when $\chi /\langle \chi \rangle \sim 10^3$, the 
corresponding intensity of energy dissipation is comparatively weak (less than $10$ the corresponding mean) -- hinting that
scalar gradient amplification may be predominantly driven by alignments
and not necessarily the strength of strain itself. 
The effect of increasing $\re$ and $Sc$ is straightforward, with 
the curves extending to larger values on both axes, 
in line with stronger intermittency at higher $\re$ and $Sc$.

On the other hand, Fig.~\ref{fig:diss}b shows the complementary
conditional expectation $\langle \chi | \epsilon \rangle$.
In this case, we observe that increasing the intensity of energy 
dissipation is accompanied by an increase in scalar dissipation, 
albeit at a slow rate. However, beyond a certain intensity of $\epsilon$, 
the curves plateau, with no increase in $\chi$ -- implying that
the most intense energy dissipation events do not lead to
more intense scalar dissipation. Once again, note 
the disparate magnitudes on the two axes, leading to the expectation
that scalar gradient amplification depends more on the alignments than on the magnitude of strain itself.

\begin{figure}
\centering
\includegraphics[height=0.32\textwidth]{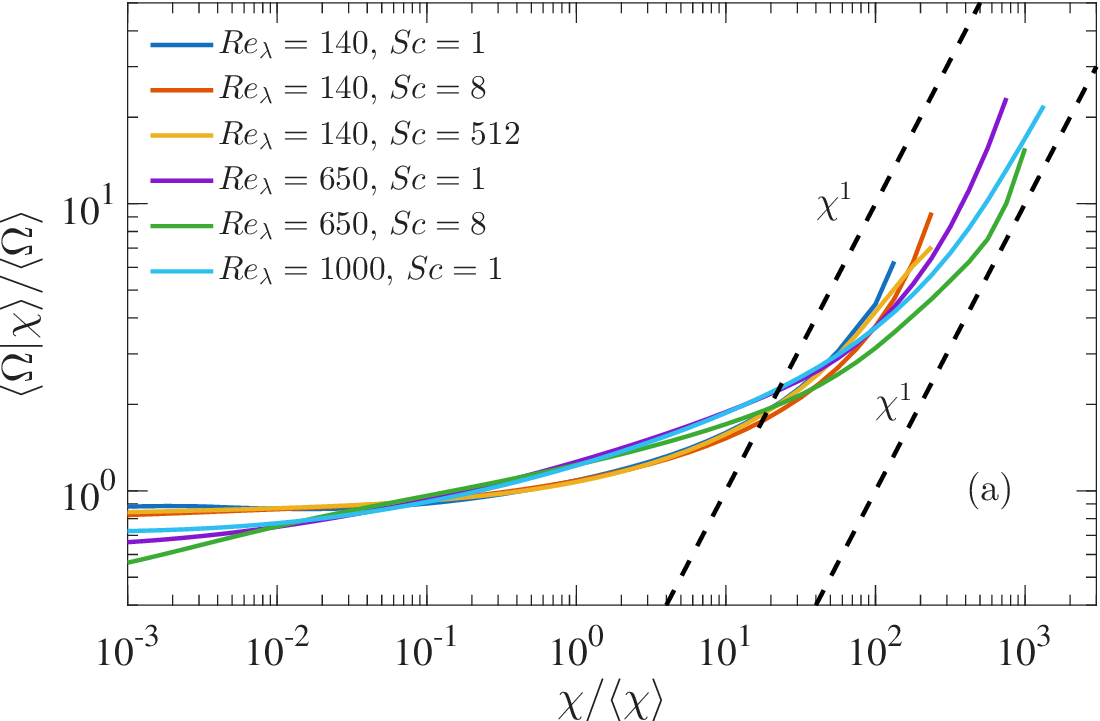} \ \  
\includegraphics[height=0.32\textwidth]{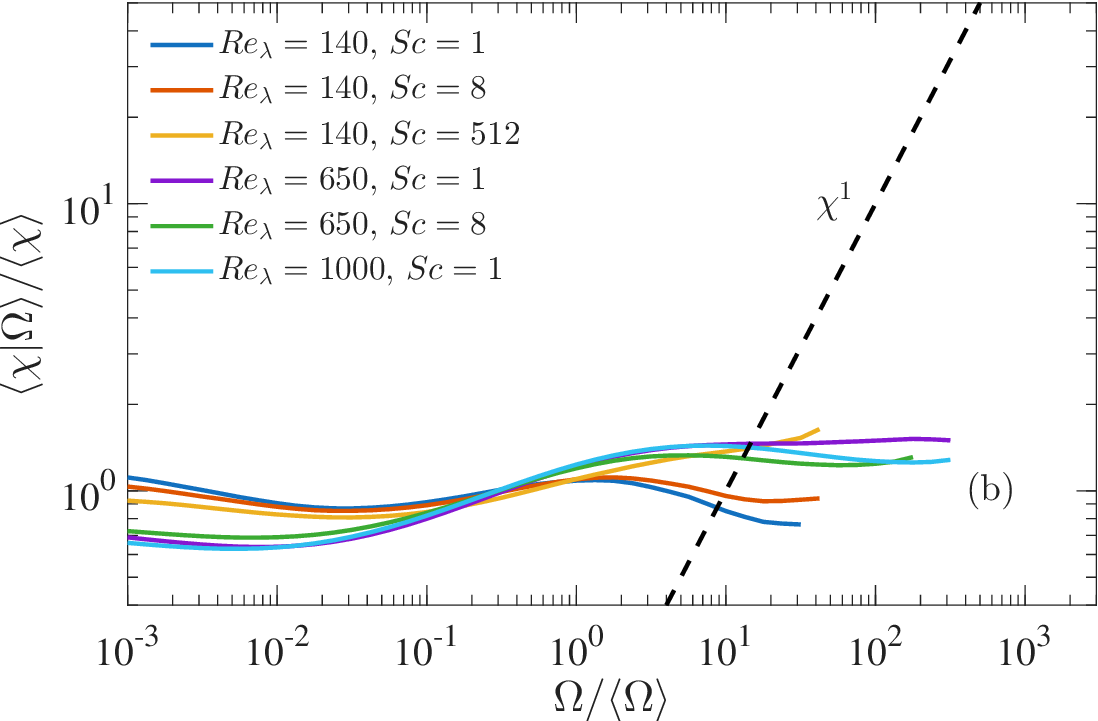}
\caption{Conditional expectations of (a) enstrophy on scalar dissipation and 
(b) scalar dissipation on enstrophy, normalized by their respective unconditional 
expectations, for various $\re$ and $Sc$.
}
\label{fig:enst}
\end{figure}

Since both $\Omega$ and 
$\epsilon $ are properties of the velocity field alone, 
these conditional statistics characterize the local velocity gradient 
structure at spatial locations where the scalar dissipation rate 
happens to be large, with no feedback from the scalar field on the 
velocity field. Figure~\ref{fig:enst} next shows the 
mutual conditional
statistics of scalar dissipation and enstrophy, with panel a
showing $\langle \Omega | \chi \rangle$
and panel b showing $\langle \chi | \Omega \rangle$ 
for various $\re$ and $Sc$. 
Interestingly, we observe that the trends in Fig.~\ref{fig:enst}a
are similar to those in Fig.~\ref{fig:diss}a, giving 
$\langle \Omega | \chi \rangle \sim \chi$ for intense $\chi$ events
(with a small proportionality factor). 
This result is perhaps to be expected given previous observations
which establish 
$\langle \Omega | \epsilon\rangle \approx \epsilon $ 
\cite{BPBY2019, BBP2020}.
Thus, even though intense scalar dissipation events are primarily
generated by strain events, the fact that intense strain also generates
intense vorticity, leads to the observed correlation between $\chi$ and $\Omega$.
This picture is indeed corroborated by the result in 
Fig.~\ref{fig:enst}b,  which shows that 
$\langle \chi | \Omega \rangle \approx \langle \Omega \rangle$, 
implying that intense enstrophy events have little to no 
statistical importance in generation of intense scalar dissipation.

\subsection{Geometry of strain tensor}
\label{subsec:cond_geom}

\begin{figure}
\centering
\includegraphics[height=0.32\textwidth]{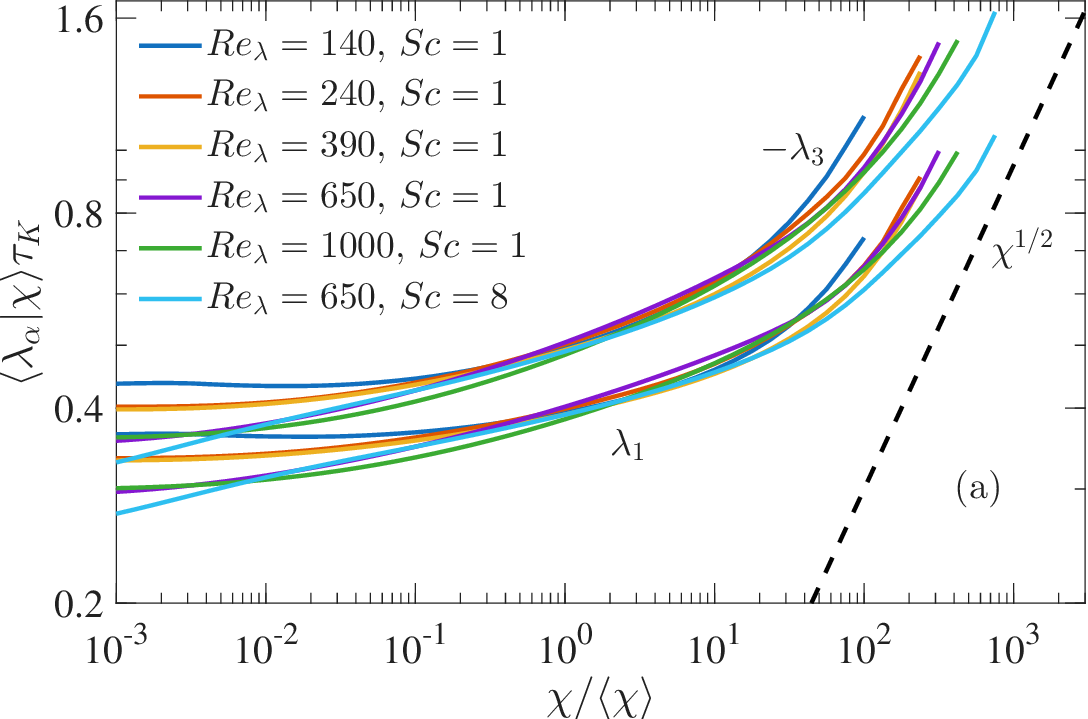} \ \ \ \ \ \   
\includegraphics[height=0.32\textwidth]{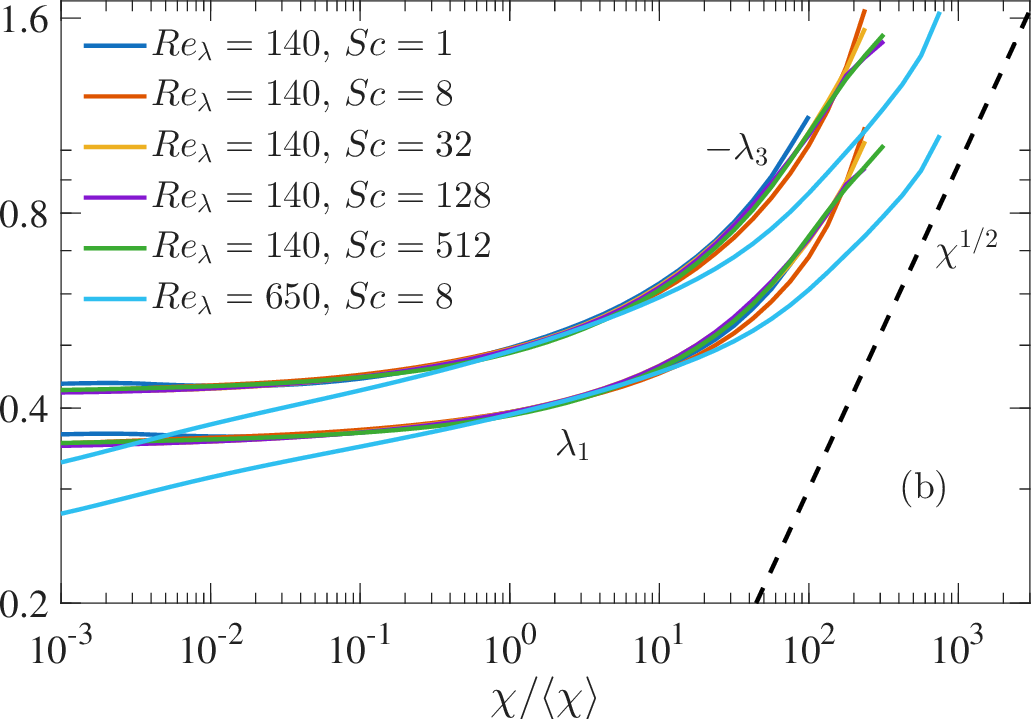}
\caption{
Conditional expectations of the extensive and compressive strain eigenvalues 
$\langle\lambda_\alpha|\chi\rangle \tau_K$, $\alpha=1$ and $3$ respectively,
for (a) various $Re_\lambda$ at 
$Sc = 1$ and (b) various $Sc$  at $Re_\lambda = 140$, 
with the $\re = 650$, $Sc = 8$ case common to both panels.
}
\label{fig:lambda}
\end{figure}

Figure~\ref{fig:lambda} shows the conditional expectations 
of most extensional and compressive strain eigenvalues: 
$\langle \lambda_1 | \chi \rangle$, 
$-\langle \lambda_3 | \chi \rangle$ (the negative sign is
used as $\lambda_3 < 0$),
to capture their relative contributions to overall 
amplitude of conditional strain.  
Panel a isolates the $\re$ dependence, while panel b
isolates the dependence on $Sc$, with the curve for
$\re=650$, $Sc=8$ providing a common reference. 
Both quantities essentially follow the behavior
of conditional energy dissipation in Fig.~\ref{fig:diss}a.
While curves Fig.~\ref{fig:lambda}a exhibit a weak dependence
on $\re$, the curves in Fig.~\ref{fig:lambda}b all collapse
as a function of $Sc$. 
Taken together, the results suggest that at 
sufficiently high $\re$ and $Sc$, 
the strain eigenvalue structure conditioned on intense scalar 
dissipation approaches a universal form.

Since the three strain eigenvalues
sum to zero (owing to incompressibility), the behavior
of $\lambda_2$ is readily deduced from the other two. Since $|\lambda_3| > |\lambda_1|$, it 
trivially follows that $\lambda_2 > 0$.
However, rather than directly analyzing $\lambda_2$, 
it is useful to consider the normalized quantity
\begin{align}
\beta = \frac{\sqrt{6}\,\lambda_2}
    {\left(\lambda_1^2 + \lambda_2^2 + 
    \lambda_3^2\right)^{1/2}},
    \label{eq:beta}
\end{align}
which is routinely used in studies of velocity gradient dynamics
and vortex stretching  \cite{lund1994, Tsi2009, buaria_jfmp}. 
It measures the relative strength of $\lambda_2$
compared to the total strain magnitude and 
is also conveniently bounded between $-1$ and $1$, 
corresponding to the limiting cases $\lambda_2 = \lambda_3$
and $\lambda_2 = \lambda_1$, respectively.

\begin{figure}
\centering
\includegraphics[height=0.32\textwidth]{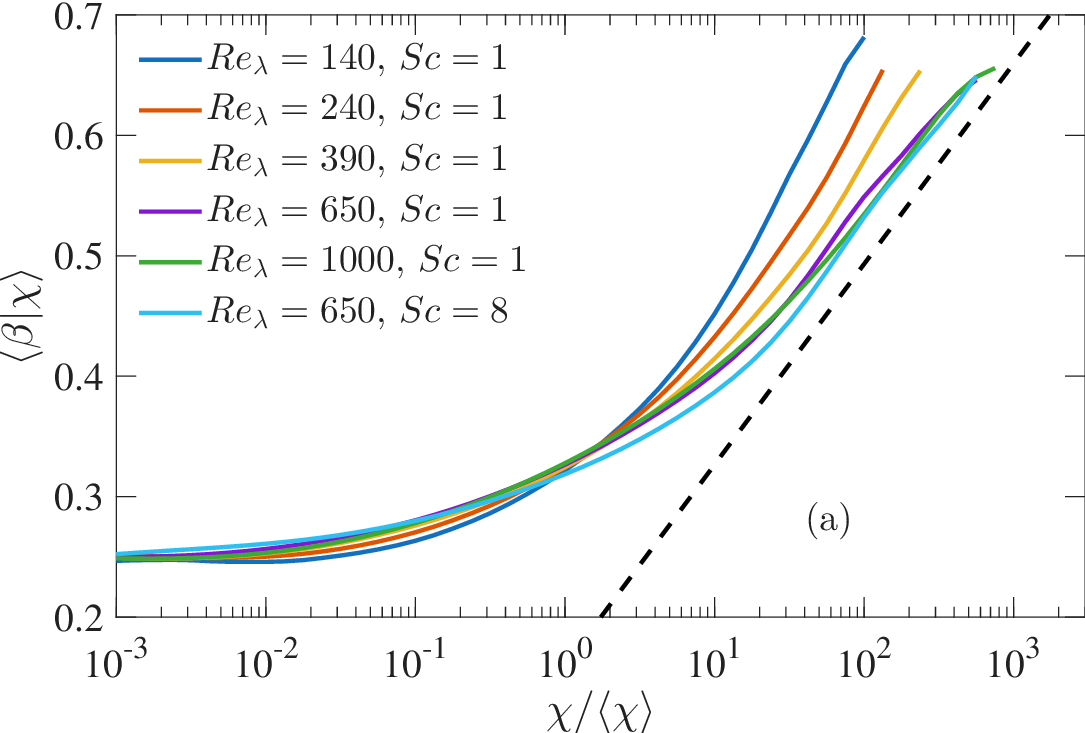} \ \ \ \ \ \   
\includegraphics[height=0.32\textwidth]{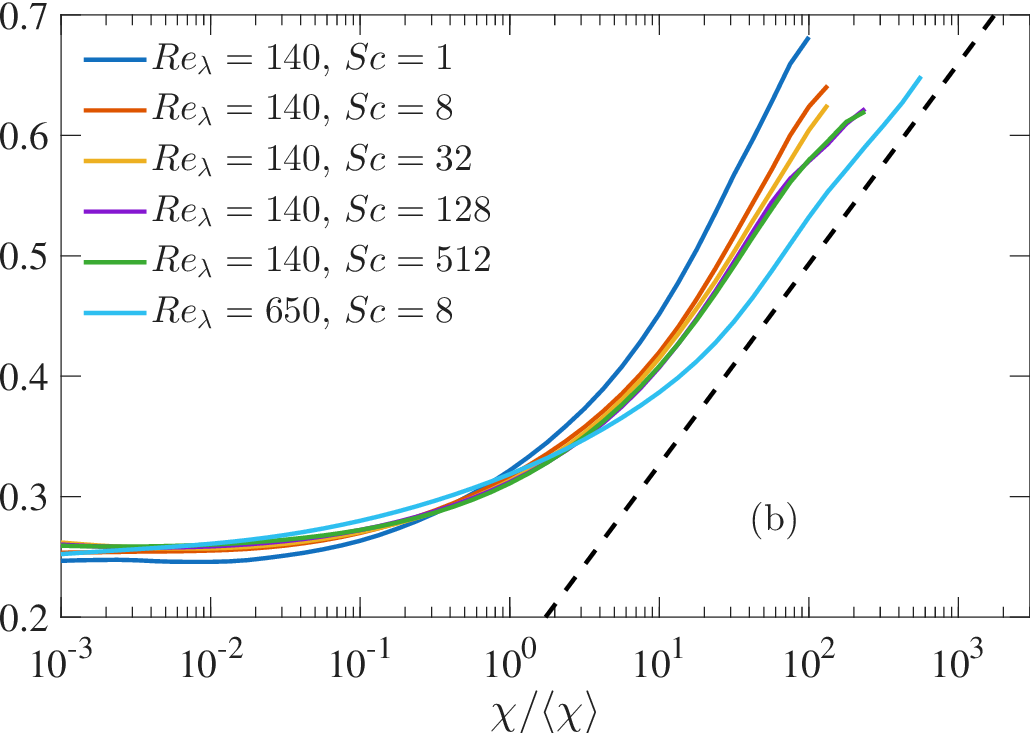}
\caption{
Conditional expectations of $\beta$ as defined by Eq.~\eqref{eq:beta} shown, 
for (a) various $Re_\lambda$ at 
$Sc = 1$ and (b) various $Sc$ at $Re_\lambda = 140$, 
with the $\re = 650$, $Sc = 8$ case common to both panels. 
The dashed line corresponds to the scaling
$\sim \tfrac{1}{6}\log_{10}(\chi / \langle \chi \rangle)$.
}
\label{fig:beta}
\end{figure}

Figure~\ref{fig:beta} shows the conditional
expectation $\langle \beta |\chi \rangle$
with panel a showing $\re$-dependence and panel b showing
$Sc$-dependence, respectively,
in direct correspondence with Fig.~\ref{fig:lambda}.
For weak events, $\chi/\langle \chi \rangle \leq 1$, 
the conditional value remains close $0.25$, which is close
to its unconditional mean \cite{BBP2020}.
As $\chi$ increases, conditional expectation
grows steadily, 
indicating that intense scalar-dissipation events preferentially 
occur in regions where the intermediate strain eigenvalue becomes 
increasingly larger relative to the total strain magnitude.
At large $\chi$, 
the conditional curves seemingly appear to get closer to 
the limiting value of unity with $\lambda_2 \to \lambda_1$. 
As before, increasing $\re$ and $Sc$ leads to an asymptotic
state, seemingly consistent with a $\log \chi$ dependence, 
as marked by the dashed line in Fig.~\ref{fig:beta}. 
Taken together with Fig.~\ref{fig:lambda}, these results
indicate that most intense scalar dissipation events are embedded
in strain events characterized by 
$-\lambda_3 > \lambda_1  \sim \lambda_2$, approaching
a state with strong compression and biaxial extension. 
This structure is precisely the one for sheet-like
organization, as also expected from prior observations.

\subsection{Alignments}
\label{subsec:cond_alignments}

\begin{figure}
\centering
\includegraphics[height=0.32\textwidth]{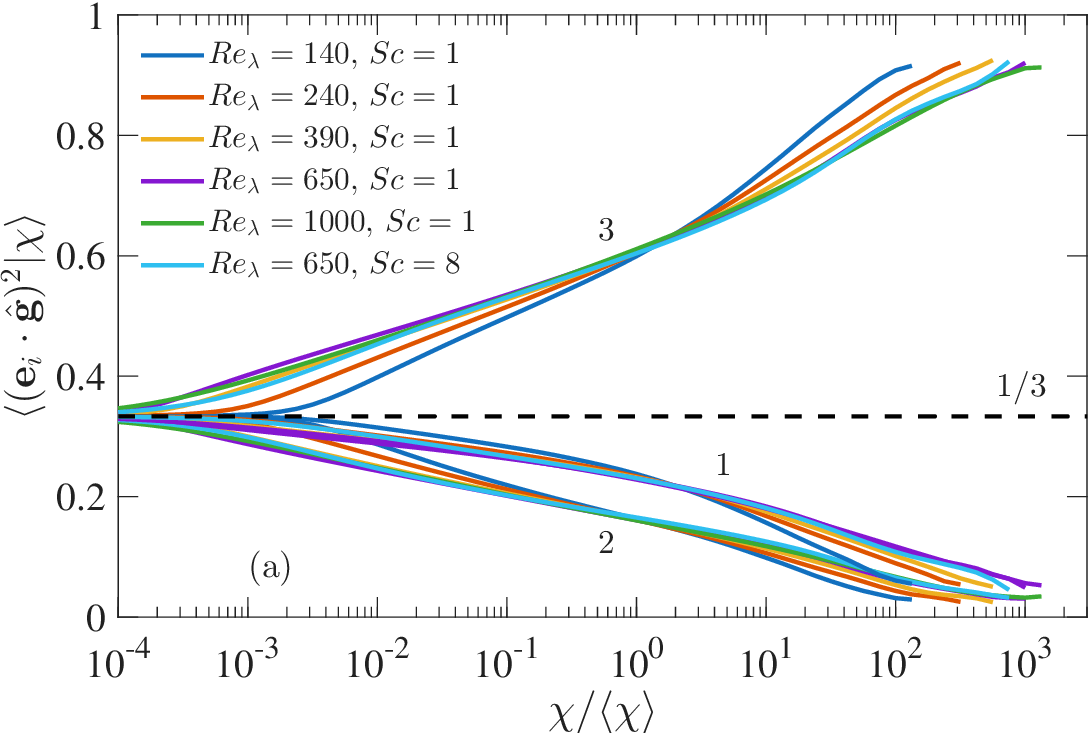} \ \ \ \ \  
\includegraphics[height=0.32\textwidth]{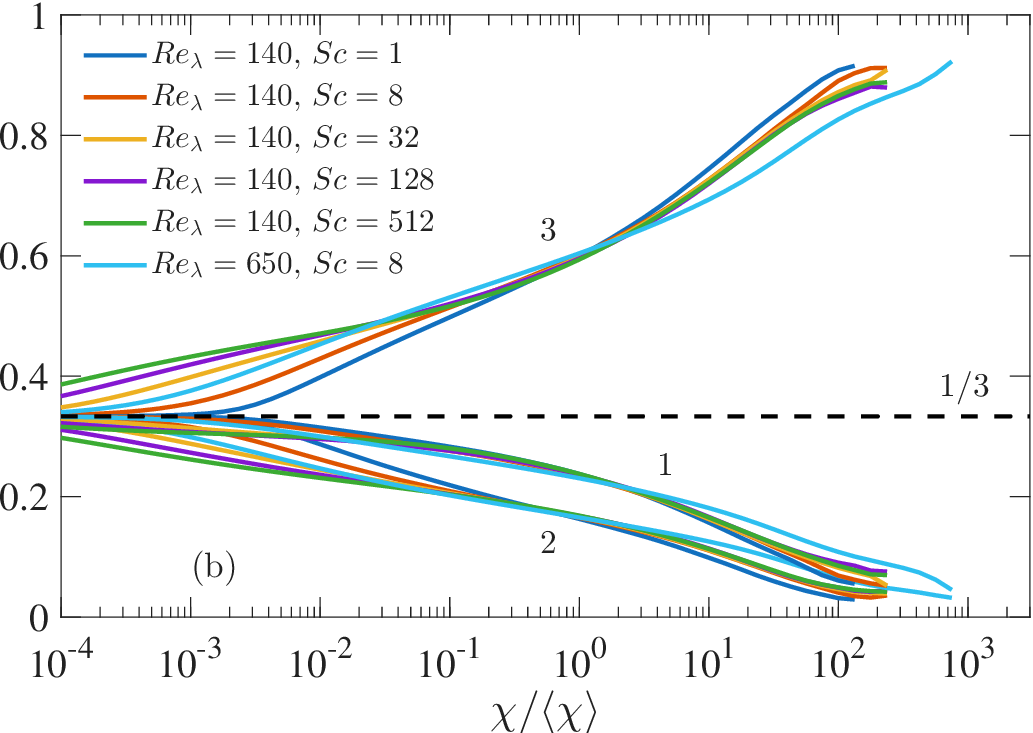}
\caption{
Conditional expectation of the second moment of the alignment cosines 
between the scalar-gradient and strain eigenvectors for 
(a) various $\re$ at $Sc=1$, and 
(b) various $Sc$ at $Re_\lambda=140$, 
with the $\re = 650$, $Sc = 8$ case common to both panels. 
The horizontal dashed line at $1/3$ marks the expectation for a
uniform distribution of the alignment cosine. 
}
\label{fig:align_ei}
\end{figure}

To complete the structural picture, we now turn to 
conditional statistics of alignments which dictate the efficacy
of amplification. 
Figure~\ref{fig:align_ei} shows the conditional second moments
of alignment cosines between scalar gradient and strain eigenvectors:
$\langle (\mathbf{e}_i \cdot \ghat)^2| \chi \rangle$, for $i=1,2,3$.
As before, panels a and b highlight 
the dependence on $\re$ and $Sc$, respectively. 
We first analyze the result in Fig.~\ref{fig:align_ei}a. 
For very weak scalar dissipation events, 
all three conditional expectations are close to $1/3$,
corresponding to a uniform distribution,
indicating that such regions are essentially structureless.
This behavior is typical of other conditional alignments, for instance,
those between vorticity and eigenvectors of strain or pressure-Hessian 
in regions on very weak dissipation and enstrophy 
\cite{BBP2020, BP2023}.
As $\chim$ increases, a pronounced 
preferential structure emerges: 
$\langle(\mathbf{e}_3\cdot\ghat)^2|\chi\rangle$ 
steadily rises toward unity, while the corresponding alignments
with $\mathbf{e}_1$ and $\mathbf{e}_2$  decrease towards zero. 
Thus, the preferential alignment of scalar gradients 
with the compressive strain direction, already observed in the unconditional 
statistics, becomes dramatically stronger when extreme 
events are considered.

The observed dependence on $\re$ in Fig.~\ref{fig:align_ei}a is weak and becomes
negligible at high $\re$, with curves for $\re \geq 650$ 
collapsing over the full range of $\chi$.
The results in Fig.~\ref{fig:align_ei}b
follow the same behavior as those in Fig.~\ref{fig:align_ei}a, 
with the dependence on $Sc$ (at fixed $\re=140$) being even weaker
when considering intense events $\chim \geq 1$ 
(although there is some visible $Sc$-dependence for weak events).
Thus, similar to results discussed in previous subsections, 
the alignment statistics also approach an asymptotic state at 
sufficiently large $\re$ and $Sc$. In this state, intense scalar dissipation
events are characterized by near-perfect alignment of scalar gradient
with the most compressive eigenvector of strain.

\begin{figure}
\centering
\includegraphics[height=0.32\textwidth]{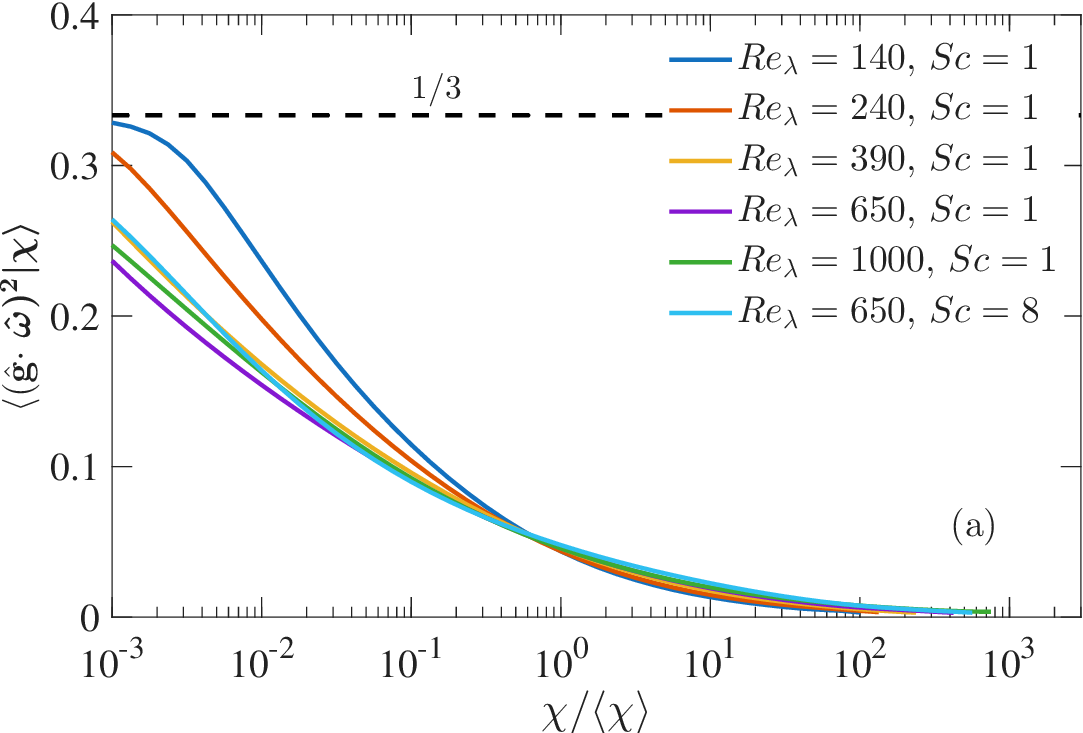} \ \ \ \ \ 
\includegraphics[height=0.32\textwidth]{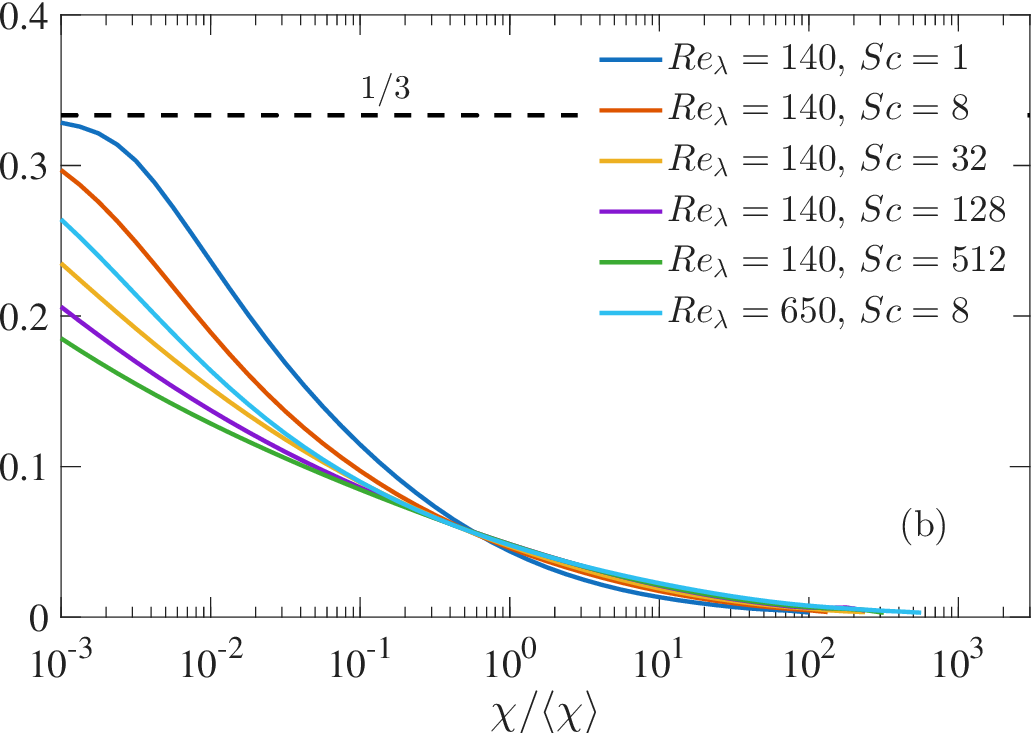}
\caption{Conditional expectation of the second moment of the 
alignment cosine between scalar-gradient and vorticity vector for 
(a) various $\re$ at $Sc=1$, and 
(b) various $Sc$ at $Re_\lambda=140$, with the $\re = 650$, $Sc = 8$ 
case common to both panels. The horizontal dashed line at $1/3$ 
marks the expectation for a
uniform distribution of the alignment cosine. 
}
\label{fig:align_gw}
\end{figure}

Finally, Fig.~\ref{fig:align_gw} shows the conditional alignment 
between scalar gradient
and vorticity: $\langle(\ghat \cdot \omhat)^2 | \chi \rangle$, 
with panels a and b once again highlighting $\re$ and $Sc$ dependencies, 
respectively. We again observe that the conditional alignments
are close to $1/3$ for very weak events, indicating a lack of structure.
However, the alignments monotonically decrease as events become increasingly
intense and essentially becomes zero for the most
intense events, implying near-perfect orthogonality between scalar 
gradients and vorticity in regions of extreme scalar dissipation.
The curves in both panels essentially collapse 
when considering intense events, suggesting no
$\re$ or $Sc$ dependence. A noticeable dependence
is observed mainly for weak events, with alignments starting to decrease
faster (as events become more intense) 
from the $1/3$ result as either $\re$ or $Sc$ is increased.
This likely reflects the increasingly sharp organization of scalar gradients 
within the local strain eigenframe as scale separation and 
intermittency become stronger.

As before, these conditional alignment statistics reveal a
remarkably coherent geometric picture of intense 
scalar dissipation. They are organized into thin
sheet-like structures, with scalar gradient preferentially aligned
with the most compressive eigenvector, both being normal to the sheet.
The two extensional strain directions lie predominantly on the plane
of the sheet, consistent with stretching in the tangential directions
and compression across the sheet thickness. Vorticity being orthogonal
to scalar gradient also lies on the sheet plane, preferentially
aligning with the intermediate strain eigenvector.
Thus intense scalar dissipation events possess a remarkably simple and 
highly organized local structure.

\subsection{Nonlinear Amplification}
\label{subsec:production_term}

\begin{figure}[ht]
\centering
\includegraphics[height=0.32\textwidth]{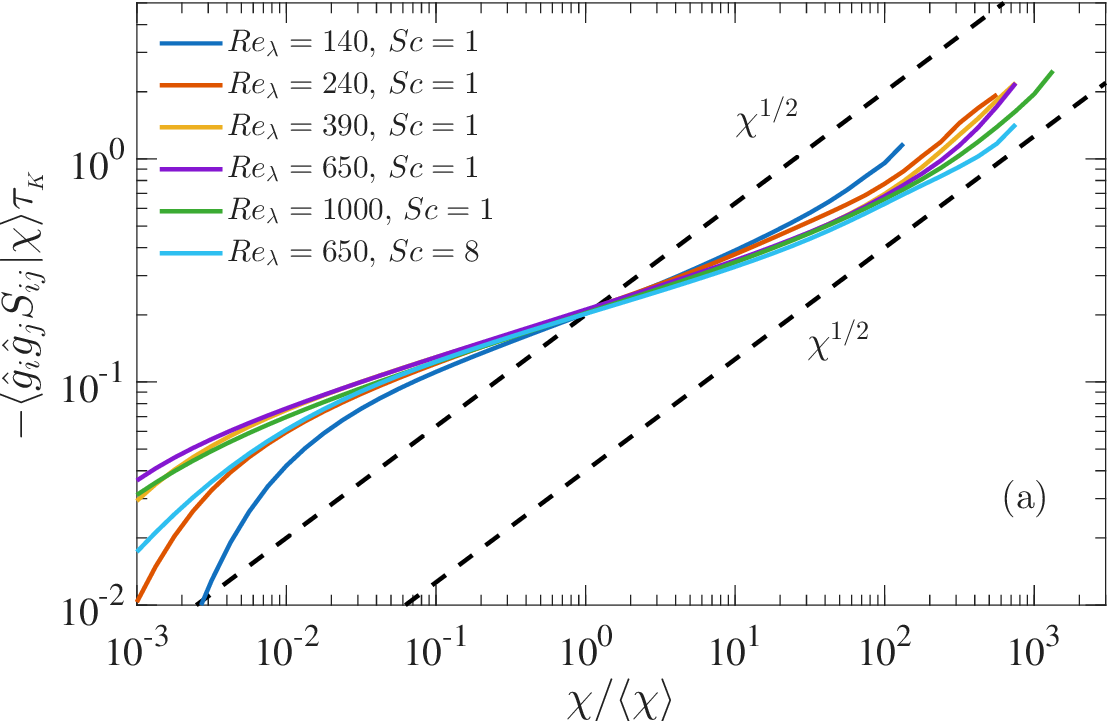} \ \ \
\includegraphics[height=0.32\textwidth]{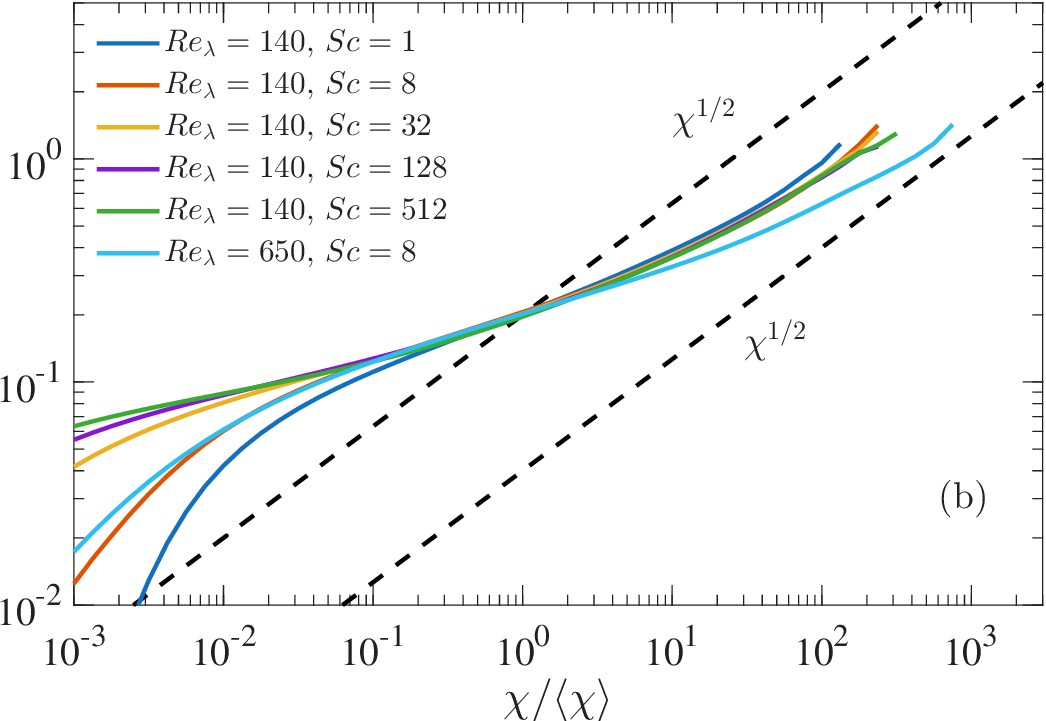} \\
\vspace{0.3cm}
\includegraphics[height=0.32\textwidth]{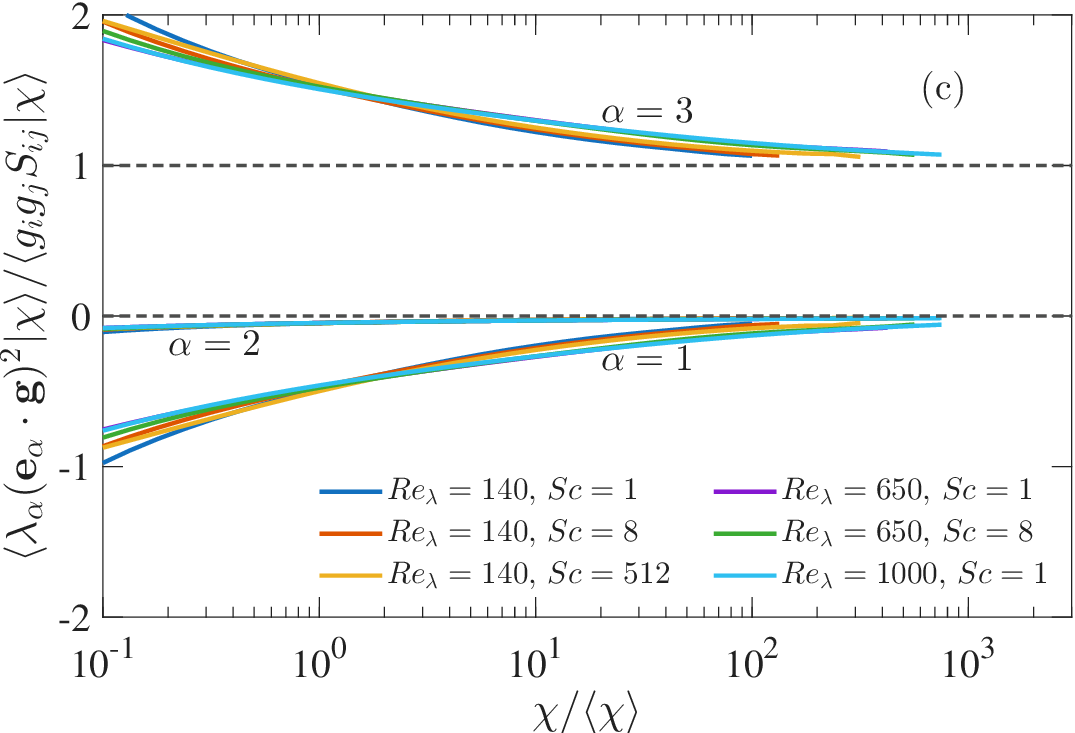}
\caption{Conditional expectation of the scalar-gradient amplification term, for (a) various $Re_\lambda$ at $Sc=1$, and (b) various $Sc$ at $\re=140$, with the $\re = 650$, $Sc = 8$ case common to both panels. 
Panel (c) shows the relative contribution from each strain eigendirection 
to the net scalar-gradient amplification.}
\label{fig:selfamp}
\end{figure}

Having characterized the conditional geometry of the
scalar gradient and the local strain fields, 
we now examine the nonlinear amplification term:
$-g_i g_j S_{ij}$. 
Since $\chi = 2D |\bg|^2$, 
the conditional expectation of this term
can simply be written as:
$ \langle g_i g_j S_{ij} | \chi \rangle 
= |\bg|^2 \, \langle \hat{g}_i \hat{g}_j S_{ij} | \chi \rangle$,
separating it into the scalar gradient magnitude and 
an {\em effective} projected strain acting on scalar gradients
to amplify it. 
Figure~\ref{fig:selfamp} shows the 
non-dimensional conditional expectation 
$-\langle \hat{g}_i \hat{g}_j S_{ij} | \chi \rangle \tau_K$,
with panels a and b isolating the $\re$ and $Sc$
dependence, respectively.
The interpretation of Fig.~\ref{fig:selfamp}
follows directly from earlier results. 
In regions 
of intense scalar dissipation, the conditional alignment
$(\mathbf{e}_3\cdot\ghat)^2 \to 1$, thus
the decomposition in Eq.~\eqref{eq:Ps_decomp}
leads to
\begin{align}
-\langle g_i g_j S_{ij} |\chi \rangle \simeq  |\lambda_3| |\bg|^2 \ ,  \qquad \chim \gg 1 \ ,
\label{eq:amp_large_chi}
\end{align}
Consequently, after removing the scalar-gradient variance, the term
$-\langle \hat{g}_i \hat{g}_j S_{ij} | \chi \rangle$
should follow the behavior of $\langle \lambda_3 |\chi\rangle$.
Indeed, this is observed in
Fig.~\ref{fig:selfamp}: the conditional expectation
increases as $\chi$ increases, seemingly approaching
a $\chi^{1/2}$ scaling at large $\chi$. 
The dependence on $\re$ and $Sc$ is also consistent with preceding
structural statistics, with
an asymptotic state reached for high $\re$ and $Sc$.

An important observation from Fig.~\ref{fig:selfamp}(a,b) is
that even for the most extreme scalar dissipation events,
the non-dimensionalized conditional expectation: 
$-\langle \hat{g}_i \hat{g}_j S_{ij} | \chi \rangle \tau_K
\simeq \mathcal{O}(1)$. 
Thus, extreme scalar gradients do not require strain rates that
are themselves extreme compared to the mean-field. 
Instead, the extreme scalar dissipation arises primarily because 
the scalar gradient increasingly aligns with the compressive strain direction
and itself becomes very large. Essentially, extreme scalar dissipation is 
controlled more by geometric organization rather than by anomalously large strain alone.

Further support for the above picture is provided by
Fig.~\ref{fig:selfamp}c, which shows the fractional contribution of 
each strain eigendirection to the total
amplification. 
The three contributions sum to unity by construction. 
For weak events, the results show a cancellation
between the most compressive and extensive eigendirections,
with the former exceeding unity, and latter being negative.
However, for the extreme events, the compressive contribution
approaches unity, with other contributions being essentially zero,
consistent with the near-perfect alignment 
of $\mathbf{g}$ with $\mathbf{e}_3$ established earlier in 
Fig.~\ref{fig:align_ei}. Remarkably, the fractional 
contributions from each eigendirection show 
virtually no dependence on $\re$ or $Sc$, 
suggesting that the relative role of each 
strain eigendirection in driving scalar gradient 
amplification is a robust geometric feature, 
independent of Reynolds and Schmidt numbers.
These results confirm that the nonlinear amplification
is almost entirely carried by the most compressive strain eigendirection,
confirming Eq.~\eqref{eq:amp_large_chi} and completing the dynamical
counterpart of the sheet-like geometric picture developed
in previous subsections.

\subsection{Role of the imposed mean-gradient}
\label{subsec:mean-gradients}

Having examined the nonlinear amplification, we now turn 
to the two mean-gradient contributions in 
Eq.~\eqref{eq:budget}, which are 
coupled with strain and rotation rate tensors, respectively.  
While their global averages are equal 
and opposite in sign, as given in Eq.~\eqref{eq:mean_grad_balance}, 
this cancellation does not hold locally or when conditioning
on scalar dissipation rate. Their conditional expectations
therefore provide a useful way to assess the memory
of imposed mean-gradient on extreme events.

\begin{figure}[ht]
\centering
\includegraphics[height=0.32\textwidth]{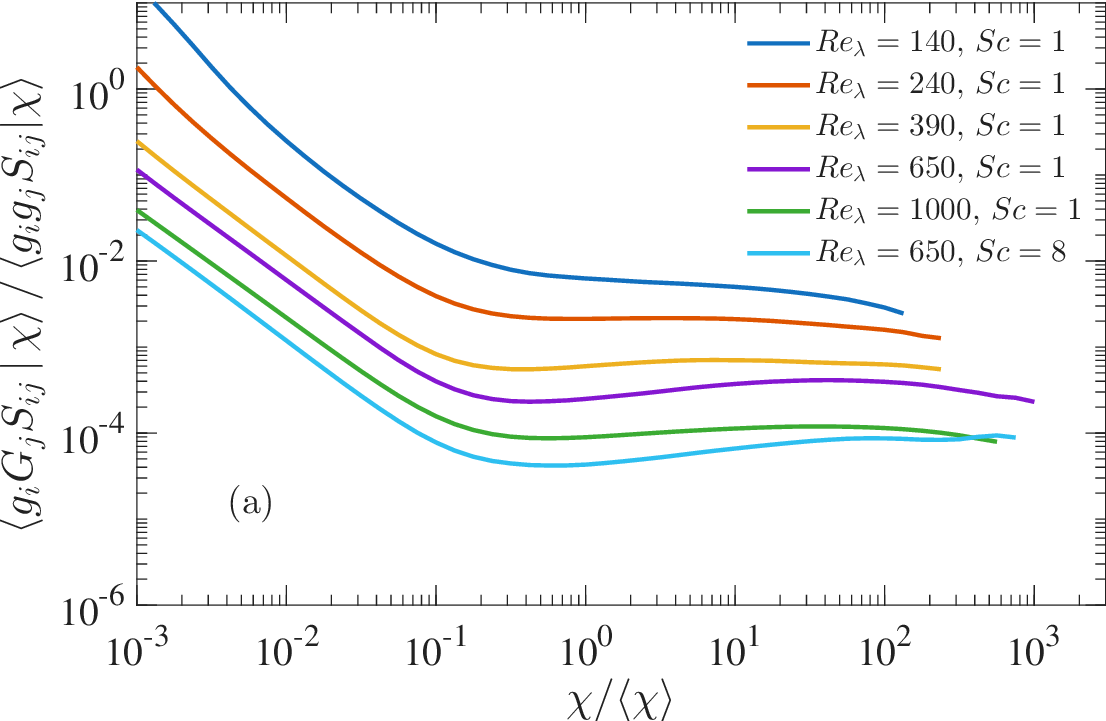} \ \ \
\includegraphics[height=0.32\textwidth]{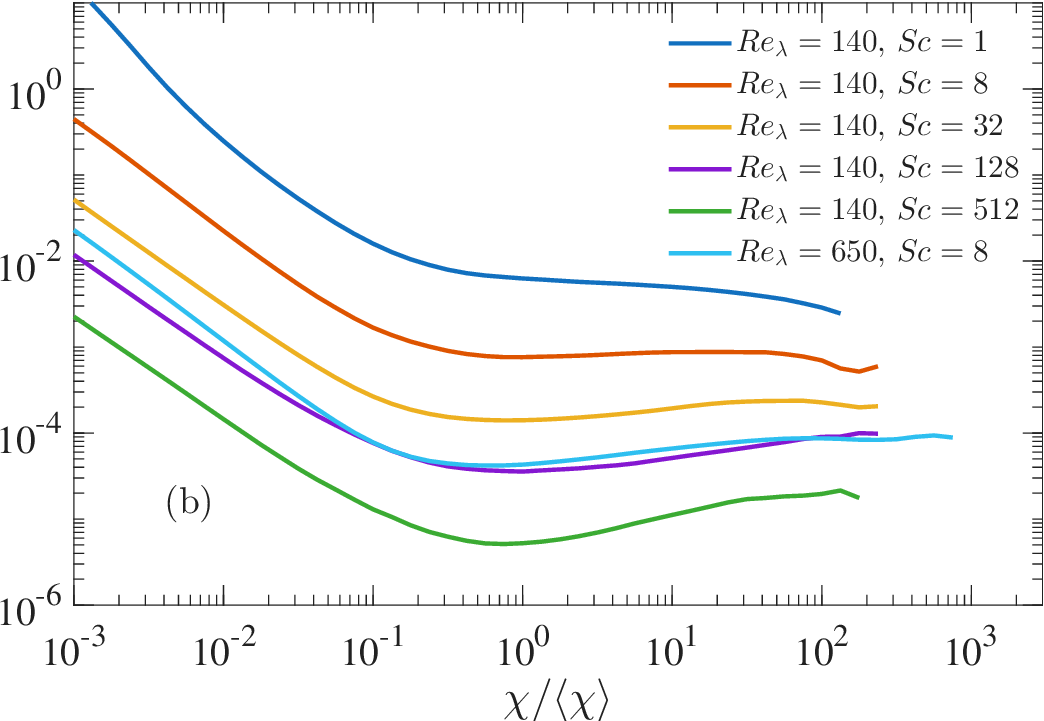} \\
\vspace{0.3cm}
\includegraphics[height=0.32\textwidth]{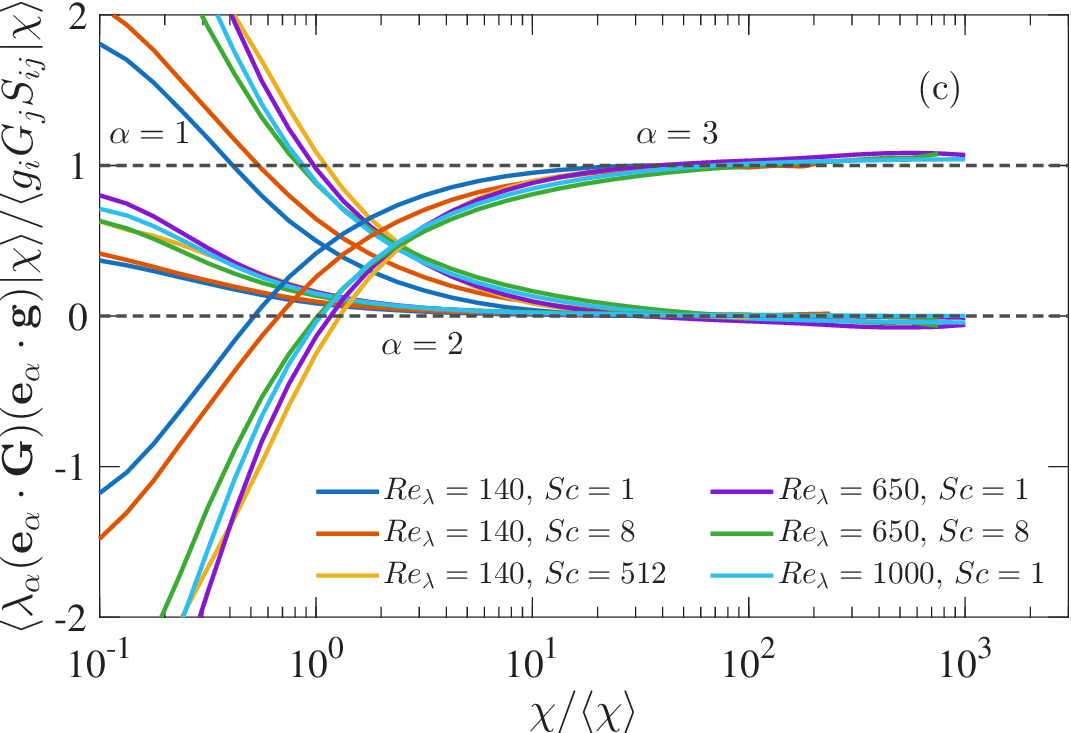}
\caption{
Conditional expectation of the mean-gradient production normalized by the 
conditional nonlinear amplification term, for (a) various $\re$ at $Sc=1$, and 
(b) various $Sc$ at
$Re_\lambda=140$, with the $\re = 650$, $Sc = 8$ case common to both panels. 
Panel (c) shows the relative contribution from each strain eigendirection 
to the mean-gradient production.
}
\label{fig:meangradient}
\end{figure}

The conditional expectation involving strain, 
$-\langle g_iG_jS_{ij}|\chi\rangle$, is shown first in
Fig.~\ref{fig:meangradient}, normalized 
by the conditional nonlinear term 
$-\langle g_ig_jS_{ij}|\chi\rangle$,
with panels a and b once again highlighting
dependence on $\re$ and $Sc$, respectively.
Several important trends are evident. 
At $\re=140$, $Sc=1$, the ratio of conditional 
contributions is large ($\approx 10$)
for weak $\chim$ events, 
but steadily decreases
as $\chim$ increases, becoming
approximately constant ($\approx 10^{-2}$) for $\chim \gtrsim 1$. 
This indicates that
the mean-gradient term plays the dominant role in amplifying
weak events, but the extreme events are predominantly generated
by nonlinear amplification. 

As either $\re$ or $Sc$ is increased (in panel a and b respectively),
the curves shift down, indicating a progressive
weakening mean-gradient contribution relative to the nonlinear term. 
This trend is consistent with the behavior of the
corresponding ratio of the unconditional averages, 
listed in Table~\ref{tab:stats_basics}.
However, as a function of $\chi$, 
the ratio of conditional averages 
are always approximately constant
for $\chim \gtrsim1$ events, indicating that once scalar
dissipation becomes sufficiently intense, 
the mean-gradient
contribution also approaches a 
scaling of $\chi^{3/2}$ in the same way
as the nonlinear amplification term.

Similar to the nonlinear term, the strain-coupled mean-gradient
term can be decomposed in the strain eigenframe as
\begin{align}
- g_i G_j S_{ij} = - \sum_{\alpha=1}^3 \lambda_\alpha \, 
(\mathbf{e_\alpha} \cdot \bg) \, (\mathbf{e_\alpha} \cdot \BG ) \ .  
\label{eq:mean_eigen_break}
\end{align}
Unlike the nonlinear term, this contribution depends not only 
on the alignment between scalar gradient and strain eigenvectors,
but also on the alignment between the imposed mean-gradient and
strain eigenvectors. 
Since $\BG$ is fixed along a Cartesian direction, its alignment with local 
strain eigenvectors is essentially random owing to isotropy of small
scales, leading to a uniform
distribution of the alignment cosines. 
However, given that $\bg$ preferentially aligns with $\mathbf{e}_3$, 
in regions where $\BG$ locally aligns with $\mathbf{e}_3$,
the mean-gradient can still generate extremely
strong scalar gradients.
Figure~\ref{fig:meangradient}c shows the  
conditional fractional contributions of the three strain eigendirections
to the mean-gradient term.
It can be readily seen that for $\chim \gtrsim1$, 
essentially only the most compressive eigendirection
contributes. 
Thus, while the mean-gradient term is overall small 
compared to the nonlinear term, it can produce
very intense scalar gradients through the same compressive
geometry, albeit rarely when $\BG$, $\bg$ and $\mathbf{e}_3$
are all parallel. 
As discussed soon, this also provides the mechanism by which 
the imposed mean-gradient breaks local isotropy.

\begin{figure}
\centering
\includegraphics[height=0.32\textwidth]{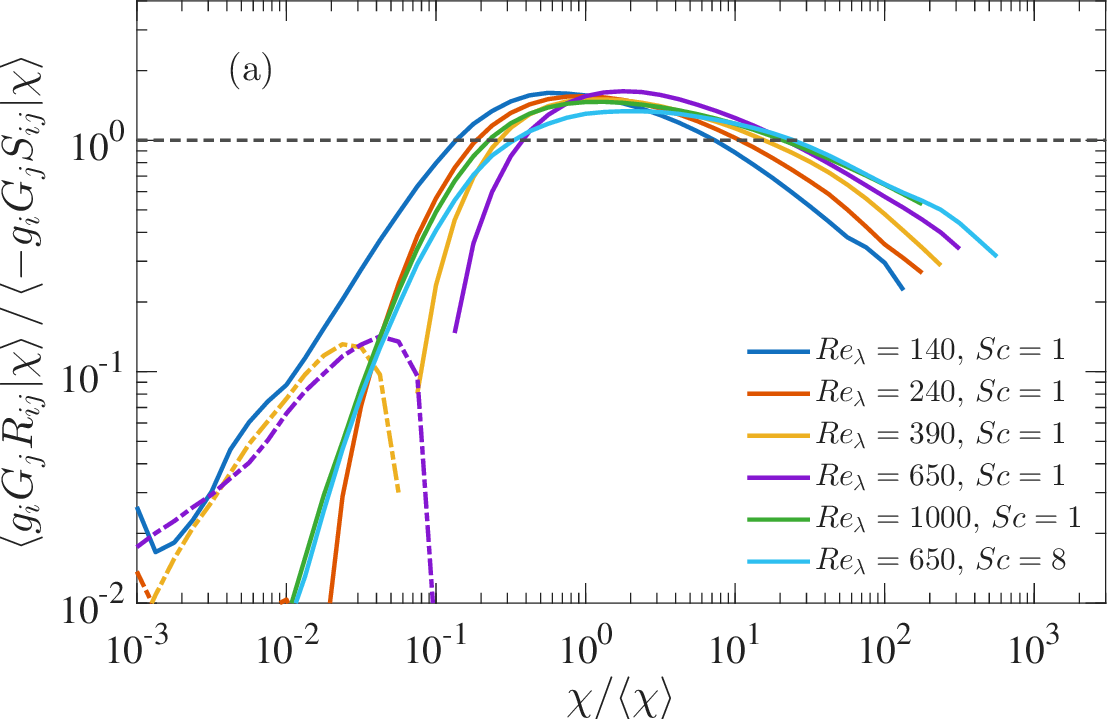} \ \ \  
\includegraphics[height=0.32\textwidth]{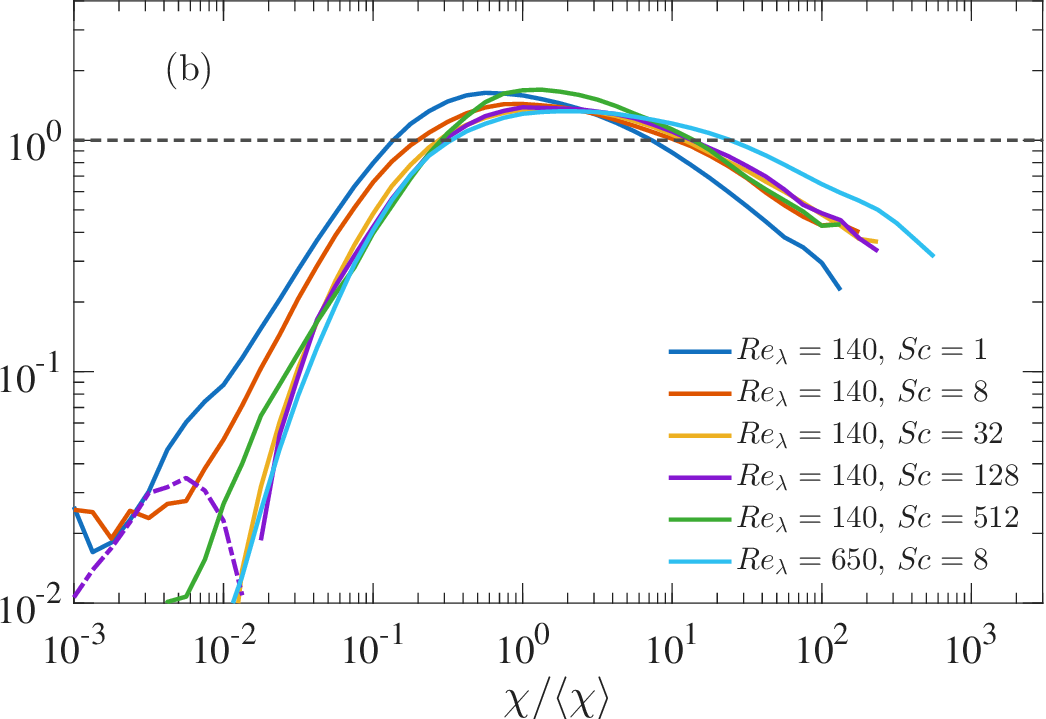}
\caption{
Ratio of conditional expectations of the rotation-coupled and strain-coupled
contributions from the mean-gradient, 
for (a) various $\re$ at $Sc=1$ and (b) various $Sc$ at $Re_\lambda=140$, 
with the $\re = 650$, $Sc = 8$ case common to both panels.  
Dash-dotted lines indicate negative contributions.
}
\label{fig:ratio_mean_rotation}
\end{figure}

We next consider the mean-gradient contribution involving
the rotation rate tensor, or equivalently vorticity,
as expressed by Eq.~\eqref{eq:rotation_identity}. 
Figure~\ref{fig:ratio_mean_rotation} 
shows the conditional expectation $\langle g_iG_jR_{ij}|\chi\rangle$,
normalized by the corresponding mean-gradient term
involving strain, $-\langle g_iG_jS_{ij}|\chi\rangle$,
since the unconditional expectations of these
two quantities are equal.
Panels a and b once again isolate the dependence
on $\re$ and $Sc$. 
Since $-\langle g_iG_jS_{ij}|\chi\rangle$ is always positive, 
the sign of the plotted ratio directly reflects the sign of 
$\langle g_iG_jR_{ij}|\chi\rangle$. Solid lines are used 
to indicate a positive contribution, acting as a production term, 
while dash-dotted lines indicate a negative contribution, 
corresponding to destruction. 
For weak $\chim$ events, the curves start off as negative 
indicating destruction of scalar dissipation, but this effect
is essentially negligible given its magnitude. 
As $\chim$ increases, the ratio changes sign, becomes positive
and grows to order unity for $\chim \gtrsim1$.
Thus, in regions of moderate and intense scalar dissipation, 
the mean-gradient contribution through vorticity
is comparable to that through strain. 
The curves in 
Fig.~\ref{fig:ratio_mean_rotation}a show only a weak
dependence on $\re$, whereas in panel b, the curves collapse with increasing $Sc$.
This suggests that the relative balance between the 
strain and vorticity-coupled mean-gradient terms also approaches an asymptotic 
state at high $\re$ and $Sc$. 
However, as discussed next, 
unlike the strain-coupled term, scalar-gradient amplification 
arising from the vorticity-coupled term
is consistent with local isotropy, with this distinction arising 
from their different
geometrical structure.

\paragraph*{Imposed mean-gradient and local isotropy:} 
It is well known that in the presence of an imposed mean-gradient,
the scalar gradient field violates local isotropy, as reflected
in order-unity skewness of $g_{\|} = \bg \cdot \hat{\BG}$, the 
scalar gradient component in the direction of the imposed
mean-gradient, even at very high Reynolds numbers 
\cite{Sreeni_1991, Yeung2002, BCSY2021a}.
This anomalous behavior is known to arise from 
characteristic ramp-cliff structures in the scalar field.
While their statistical behavior has been extensively 
characterized, the dynamical route by which the imposed
mean-gradient produces this behavior 
remains unclear. 
Since the mean-gradient amplification term
interacts with both strain and vorticity, it is natural to 
ask which one of these is responsible for transmitting
the large-scale anisotropy to small scales.

For the strain-coupled term, 
we saw that the dominant contribution
is associated with the most compressive
eigendirection of strain. 
Since $\bg$ is preferentially aligned
with $\mathbf{e}_3$, the decomposition
in Eq.~\eqref{eq:mean_eigen_break} 
gives
$-g_i G_j S_{ij} \approx  -  \lambda_3 \, 
(\mathbf{e_3} \cdot \bg) \, (\mathbf{e_3} \cdot \BG )$   
which leads to amplification when
$ (\mathbf{e_3} \cdot \bg) \, (\mathbf{e_3} \cdot \BG ) > 0$
since $\lambda_3 < 0$.  
This condition essentially corresponds to 
preferential amplification of the scalar gradient
component whose projection along the imposed
mean-gradient is positive, i.e., $g_{\|} > 0$.
In contrast, negative values of $g_{\|}$
correspond to $-g_i G_j S_{ij} < 0$ 
and are depleted.
Thus, the strain-coupled mean-gradient term
naturally introduces a sign bias,
providing a dynamical explanation
of how fronts with $g_{\|} > 0$
are generated at a higher rate in the flow, leading to violation
of local isotropy.

In contrast, the vorticity-coupled term has a different 
geometrical structure. 
Following Eq.~\eqref{eq:rotation_identity}, the term 
involves the projection  
$\bg \times \ww$ along $\BG$, rather than the projection
of $\bg$ along $\BG$. 
Since $\bg$ and $\ww$ are preferentially orthogonal
and respectively aligned with 
$\mathbf{e}_3$ and  $\mathbf{e}_2$, 
it follows that $\bg \times \ww$
is preferentially aligned with $\mathbf{e}_1$. 
Thus, the vorticity-coupled term leads to amplification
when $\mathbf{e}_1$ is aligned with $\BG$.
However, this condition is not tied to the sign of any component
of $\bg$ and amplifies both positive and negative gradients equally.

\begin{figure}
\centering
\includegraphics[width=0.9\textwidth]{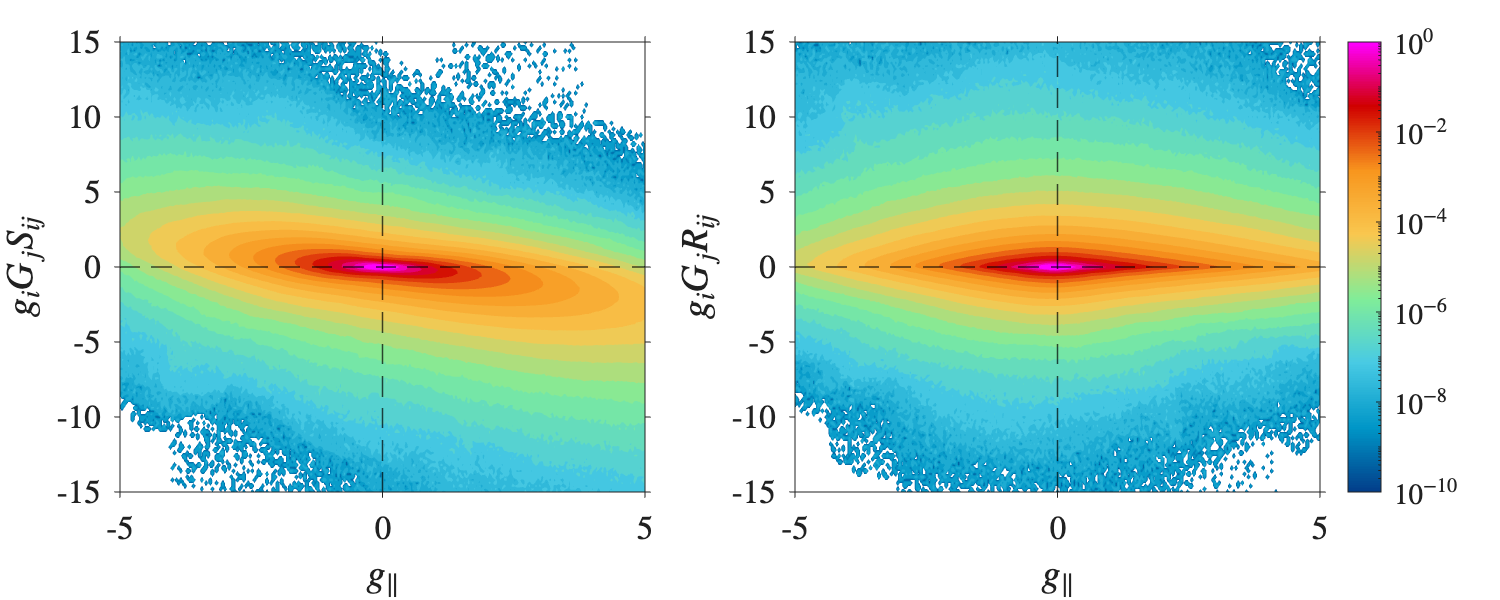}\\
\includegraphics[width=0.9\textwidth]{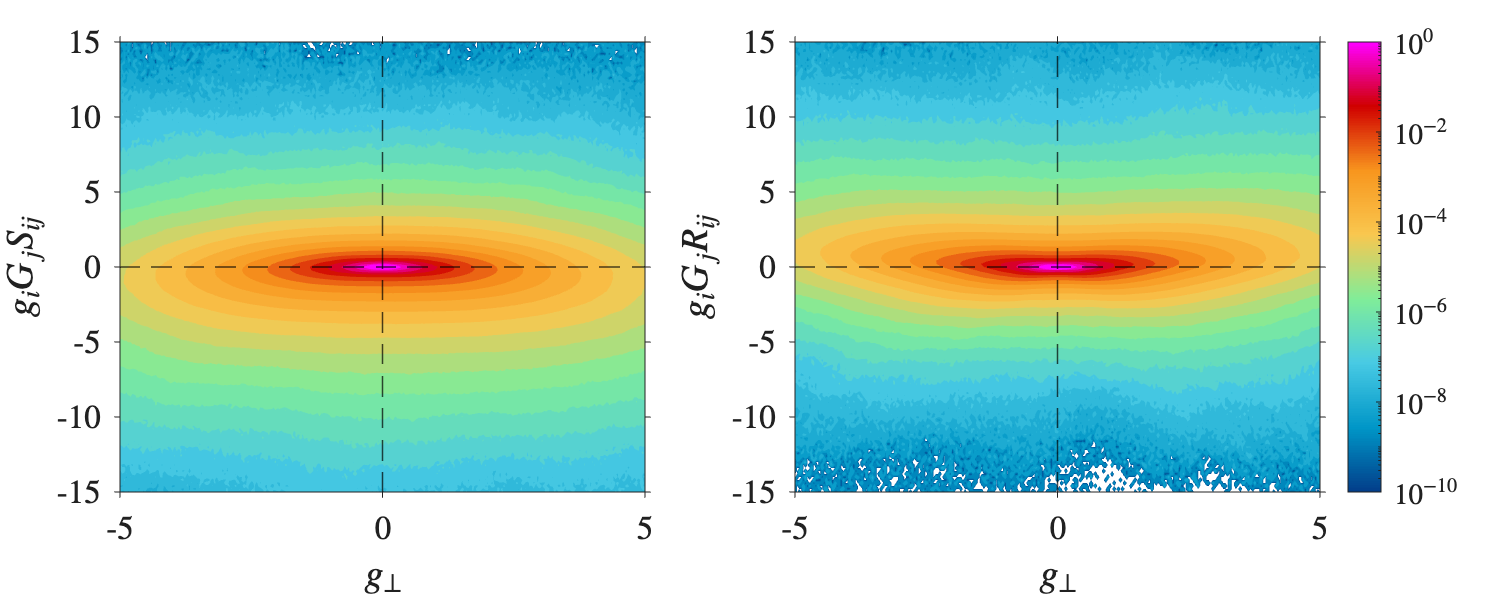}
\caption{
Joint probability density functions (p.d.f.s) of 
$g_\parallel$ (top row),  $g_\perp$ (bottom row), the 
scalar gradient components parallel and perpendicular to the 
imposed mean-gradient $\mathbf{G}$, with the amplification terms
$g_iG_jS_{ij}$ (left column), $g_iG_jR_{ij}$ (right column), 
at $Re_\lambda = 140$, $Sc = 1$.
The scalar gradients is normalized by its $L^2$-norm: $\langle g_ig_i\rangle^{1/2}$
and the strain and rotation tensors by the Kolmogorov time
scale $\tau_K$.
The corresponding result for $\re=140$, $Sc=512$ is shown in 
Fig.~\ref{fig:joint_pdf_140_512} in Appendix~\ref{app:joint_pdfs}. 
}
\label{fig:joint_pdfs}
\end{figure}

Figure~\ref{fig:joint_pdfs} tests these interpretations
directly by examining the joint probability distributions
between scalar gradient components and mean-gradient amplification
terms at $\re=140$, $Sc=1$. 
The top row shows the joint distributions for $g_{\|}$ 
and mean-gradient terms, while the bottom row
shows $g_\perp$, the component of $\bg$ which is perpendicular
to $\BG$. 
The first panel in each row show the joint
distributions with $g_i G_j S_ij$, whereas the second panel
with $g_i G_j R_ij$.
Focusing on plots in the top row, 
a clear distinction can be made.
The distributions are asymmetric around $g_{\|} =0$,
with positive value more likely as expected \cite{Yeung2002, BCSY2021a}.
However, for the strain-coupled term, 
the distribution is strongly 
asymmetric around $g_i G_j S_{ij} = 0$, 
with $g_i G_j S_{ij} < 0$ is more likely to occur with 
$g_{\|} > 0$, and  $g_i G_j S_{ij} > 0$ 
with $g_{\|} < 0$.
Thus, the strain-coupled mean-gradient term 
selectively amplifies $g_{\|} >0$, while attenuating
$g_{\|} < 0$. 

In contrast, the vorticity-coupled term is essentially symmetric
around $g_i G_j R_ij = 0$, implying no preferential amplification
of positive or negative $g_{\|}$. 
Both the corresponding distributions with $g_\perp$ shown
in the bottom row are fully symmetric all around, as expected.
Thus, as reasoned earlier, the violation of local isotropy
due to the imposed mean-gradient arises from its
coupling with strain. 
Similar result as Figure~\ref{fig:joint_pdfs}
is shown for $\re=140$, $Sc=512$ in Appendix~\ref{app:joint_pdfs}.
In this case, the distributions become more symmetric
around $g_{\|}=0$ as expected from approach to isotropy 
with increasing $Sc$ \cite{BCSY2021a}.
However, the distribution for the strain-coupled term
still breaks symmetry. This implies that the mean-gradient
term always transfers large-scale anisotropy to the smallest
scales, but with increasing $Sc$ the relative importance
of the mean-gradient term with respect to the nonlinear
term reduces dramatically, which indeed was also observed in
Fig.~\ref{fig:meangradient}.

\subsection{Role of diffusive terms}
\label{subsec:diffusive_terms}

In the previous subsections, we examined the dynamics
of all the terms containing velocity gradients 
which lead to the production of scalar dissipation.
In contrast, the two diffusive terms in Eq.~\eqref{eq:budget} regulate
the destruction of scalar dissipation. 
The scalar Hessian term $ D ||\nabla \bg||^2$ is positive definite 
and therefore always contributes to the destruction of scalar dissipation. 
In contrast, the Laplacian term 
$-D\nabla^2 (g_i g_i)$, 
is on average zero (from homogeneity) and hence does not contribute to the
mean-budget. Rather, it acts as diffusion term leading to 
a spatial redistribution of scalar dissipation, and thus, can either 
produce or destroy scalar dissipation depending on the local strength
of scalar dissipation. 

\begin{figure}[ht]
\centering
\includegraphics[height=0.31\textwidth]{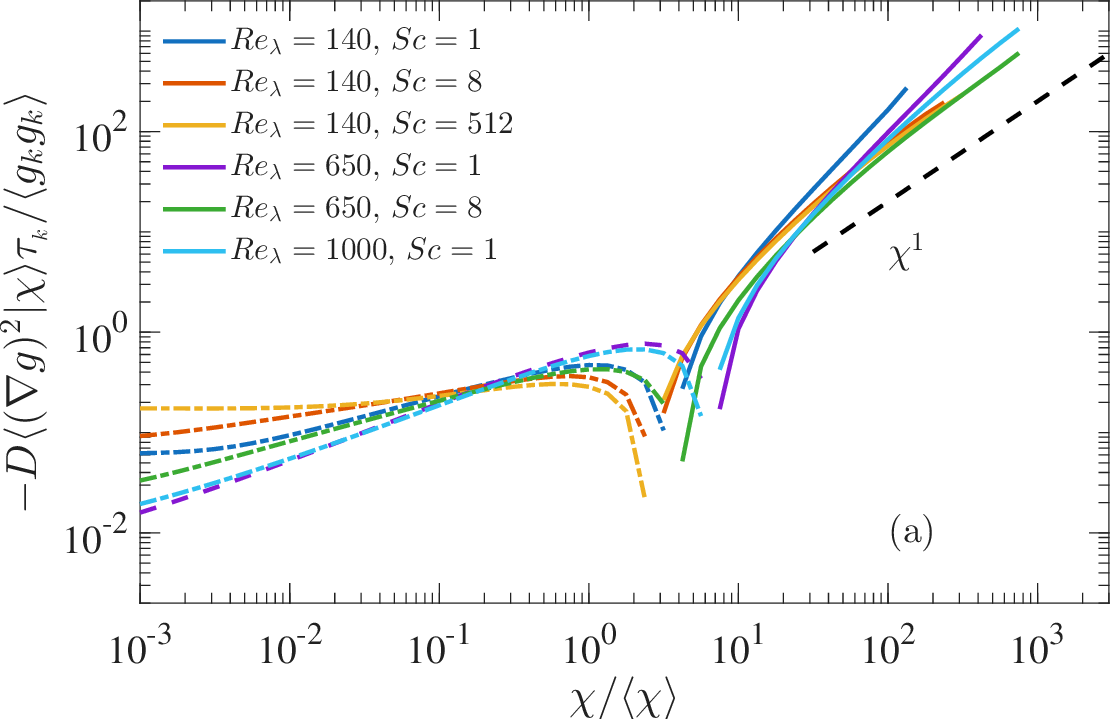} \ \ \ 
\includegraphics[height=0.31\textwidth]{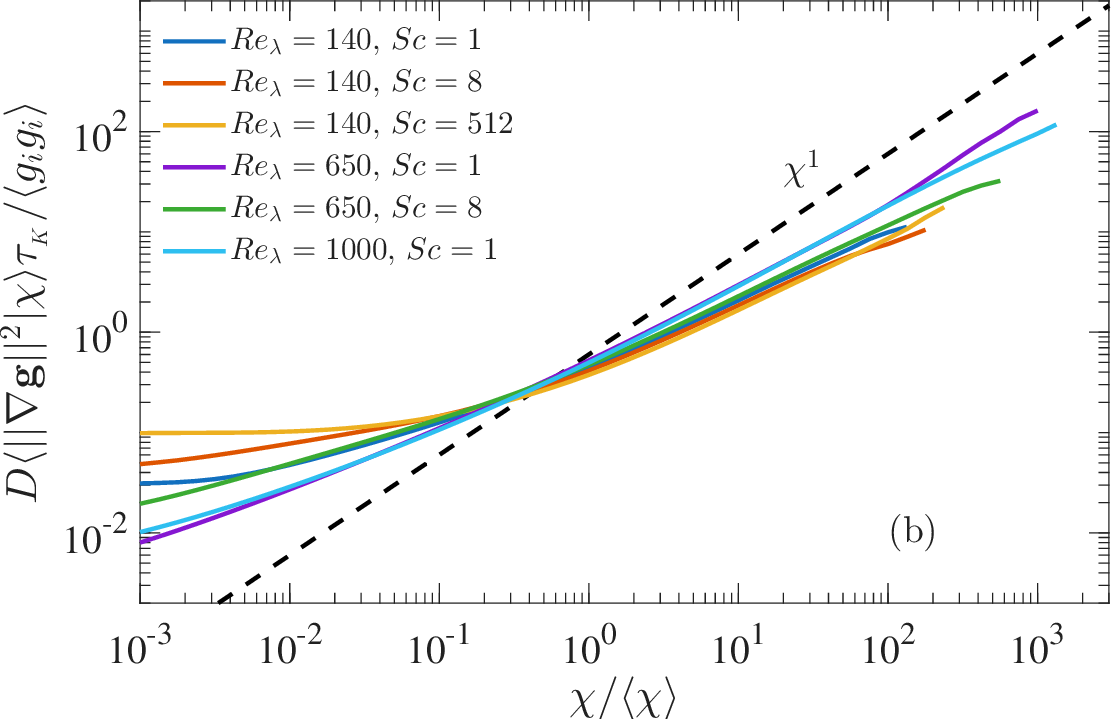}
\caption{
Conditional expectations of (a) the diffusive Laplacian term, and (b) 
the Hessian destruction term 
for different $Re_\lambda$ and $Sc$ values. 
For the diffusive term in panel a, solid lines denote
positive values, while dot-dashed lines denote negative values. 
}
\label{fig:diffusion_dissipation}
\end{figure}

Figure~\ref{fig:diffusion_dissipation} shows the conditional expectations
of these two terms. Panel a shows the the Laplacian term $-D\nabla^2 (g_i g_i)$,  
and panel b shows the Hessian term $ D ||\nabla \bg||^2$, both appropriately
non-dimensionalized. Since the former is zero on average, 
and the figure utilizes log-scales, the positive values are represented
by solid lines, while the magnitude of the negative values is shown using 
dash-dotted lines. 
The curves for the Laplacian term in panel a
naturally separate into weak and intense scalar dissipation
events, with  negative contribution for $\chim \lesssim 1$ 
and positive contribution for $\chim \gtrsim 1$, consistent with 
a diffusive redistribution.

The Hessian destruction term $ D ||\nabla \bg||^2$, which is always
positive, is shown in
Fig.~\ref{fig:diffusion_dissipation}b.  
Interestingly, we observe that for weak events, the curves are virtually identical to those
for the diffusive Laplacian term in Fig.~\ref{fig:diffusion_dissipation}a,
pointing to cancellation between the two, with the net contribution
of diffusive terms in weak scalar dissipation regions being essentially
negligible. 
For intense events, both the Laplacian and Hessian 
terms rapidly increase with the conditioning
value. However,  the magnitude of the 
Laplacian term is larger than that of the
Hessian term, indicating that the arrest of extreme scalar dissipation
is dominated by diffusive transport, rather than by purely local 
destruction. In fact, this behavior closely parallels the 
viscous terms in enstrophy dynamics, 
where viscous diffusion redistributes enstrophy from intense vortex
tubes to surrounding regions, while the viscous destruction
arrests enstrophy locally \cite{BBP2020}. 
Thus, even though the amplification dynamics of scalar dissipation
and enstrophy are fundamentally different, their destruction
mechanisms closely mirror each other.
Even the $\chi^1$ scaling observed in 
Fig.~\ref{fig:diffusion_dissipation}a can be deduced
by recognizing that the Laplacian 
$D \nabla^2 (g_i g_i)$  scales as $ D g_i g_i / \eta_B^2$,
which when non-dimensionalized by $\tau_K / \langle g_i g_i\rangle$,
exactly gives $\chim$.

\section{Conclusions}
\label{sec:conc}

In this work, we have investigated the structure
and dynamics of scalar gradients in turbulent mixing, 
with particular emphasis on the mechanisms responsible
for generation of intense scalar dissipation events.
To that end, we utilized a massive database of well-resolved 
direct numerical
simulations (DNS) of stationary isotropic turbulence, 
together with passive scalars driven by a uniform mean-gradient.
The Taylor-scale Reynolds number $\re$ is in the range
$140-1000$ on grid sizes going up to
$8192^3$, and the Schmidt number $Sc$ is in the range $1-512$.
By analyzing various unconditional and conditional
statistics derived from the budget equation
of scalar dissipation, we determine how scalar gradients
are amplified, their geometric organization, role of
imposed mean-gradient and molecular diffusion, and
also how these mechanisms depend on $\re$ and $Sc$. 

The unconditional statistics reaffirm, and extend
to higher $\re$ and $Sc$, the known picture
of scalar-gradient amplification by strain. 
Scalar gradients preferentially align with the most
compressive strain eigenvector $\mathbf{e}_3$, and remain orthogonal
to vorticity, with these correlations being essentially
independent of $\re$ and $Sc$. 
The mean nonlinear amplification 
$-\langle g_i g_j S_{ij} \rangle$, which can be related
to mixed velocity-scalar derivative skewness,
is dominated
by the compressive eigendirection, 
while the most extensional direction provides a weak 
attenuating contribution and the intermediate direction 
contributes negligibly.

The conditional statistics reveal that this average picture
becomes much sharper in regions of intense scalar dissipation.
The scalar gradients approach near-perfect alignment with 
$\mathbf{e}_3$ for extreme events, 
while becoming orthogonal 
to other strain eigenvectors and to vorticity.
The strain eigenvalue geometry suggests $\lambda_2 \simeq \lambda_1$.
These results, also supported by visualizations, essentially
show that intense scalar dissipation
is organized into thin sheet-like structures
embedded in the shear layers between vortex tubes.
The amplification term for extreme events essentially reduces
to $- g_i g_j S_{ij} \simeq |\lambda_3| |\bg|^2$. 
However, the effective projected strain
$- \hat{g}_i \hat{g}_j S_{ij}$ when normalized
by the Kolmogorov time scale $\tau_K$, is 
order unity, revealing that 
intense scalar dissipation events do not 
necessarily arise from intense strain, but predominantly 
due to near-optimal compressive alignment. 
This suggests that the scalar-variance cascade may be
more efficient than the kinetic-energy cascade; 
analysis analogous to that recently carried out for 
the velocity field \cite{ballouz2018},
could perhaps shed further light on the  mechanisms underlying this 
enhanced transfer.

The contribution from imposed mean-gradient terms is much weaker 
than the nonlinear amplification term in the overall 
budget and is essentially 
negligible at high $\re$ and $Sc$. 
Nevertheless, they still transfer residual anisotropy
directly from large to smallest scales.
Specifically, we reason that the coupling of the mean-gradient 
with strain leads to selective amplification
of the positive values of 
$g_{\}} = \bg \cdot \hat{\BG}$, the scalar gradient component 
projected along the mean-gradient, while attenuating
the negative values. 
This produces the persistent
skewness of $g_{\|}$ associated with ramp-cliff structures,
a feature widely observed in scalar turbulence 
\cite{Sreeni_1991, Shraiman2000, BCSY2021a}. 
By contrast, the coupling of the mean gradient with 
vorticity amplifies scalar gradients without 
producing any sign asymmetry.

Finally, the diffusive terms reveal how intense scalar dissipation
events are arrested. In the mean-budget, 
the positive-definite scalar Hessian term
provides the local destruction that balances net production, 
whereas the Laplacian term  has zero mean
by homogeneity and therefore acts only to redistribute 
scalar dissipation in space. 
However, when conditioning on intense scalar dissipation, 
this redistribution term becomes the dominant destructive 
contribution. It removes scalar dissipation from intense structures and 
transports it toward weaker regions, while the local Hessian destruction 
remains subdominant. 
This aspect is essentially similar to the 
destruction of enstrophy or energy dissipation in their 
respective budgets, where viscous diffusion plays a central 
role in depleting the most intense events \cite{BBP2020, BPB2022}.

Nearly all the statistics reported here are either independent
of $\re$ and $Sc$ from the outset, or become so at 
high $\re$ and $Sc$, revealing a remarkably universal
structure of the small-scales of the scalar field.
Since the amplification of scalar gradients 
is driven by the local velocity gradients, this universality
is likely rooted in the organization of the velocity
gradients themselves. Indeed, recent works
have shown that velocity gradient statistics exhibit
a universal structure across different turbulent
flows \cite{BP:2025}, suggesting that the scalar gradient dynamics
identified here should also extend
beyond the present setting of stationary isotropic
turbulence. A natural extension of this work is therefore
to examine and compare scalar-gradient structure in different
turbulent flows; this will be reported in future work.

\begin{acknowledgements}
\paragraph*{Acknowledgements}
The authors gratefully acknowledge the Gauss Centre for Supercomputing 
e.V. (www.gauss-center.eu) for providing time on the supercomputer
JUWELS at J\"ulich Supercomputing Centre (JSC).
We also acknowledge the Texas Advanced Computing Center (TACC) 
at UT Austin (www.tacc.utexas.edu) for providing computational 
resources that have contributed to the research results reported 
within this paper.  
The high Schmidt number simulations
using the hybrid approach were performed together 
with Matthew P. Clay and P. K. Yeung using computational resources at 
the Oak Ridge Leadership Computing Facility (OLCF), under 2017
and 2018 INCITE Awards.
\end{acknowledgements}

\appendix

\section{Derivation of Eq.~\eqref{eq:mean_grad_balance}}
\label{app:identity}

Using $g_i = \partial \theta / \partial x_i$ and noting that $G_j$ is constant
it follows 
\begin{align}
\langle g_i G_j A_{ij} \rangle 
&= G_j \left \langle \frac{\partial \theta}{\partial x_i}\,
\frac{\partial u_i}{\partial x_j} \right \rangle
\label{eq:app1}\\[6pt]
&= G_j \left \langle \frac{\partial}{\partial x_i}\!\left(
\theta\,\frac{\partial u_i}{\partial x_j}\right) \right \rangle
- G_j \left \langle \theta\,
\frac{\partial^2 u_i}{\partial x_i\,\partial x_j} \right \rangle
\label{eq:app2}\\[6pt]
&= 0,
\label{eq:app3}
\end{align}
where the first term is zero from statistical homogeneity and the 
second term from incompressibility.
Since the velocity gradient tensor can be decomposed as 
$A_{ij} = S_{ij} + R_{ij}$, it follows that
\begin{align}
\langle g_i G_j S_{ij} \rangle + \langle g_i G_j R_{ij} \rangle &= 0,
\end{align}
leading to the result in Eq.~\eqref{eq:mean_grad_balance}.

\section{Derivation of Eq.~\eqref{eq:Mijkl_final}}
\label{app:isotropic_tensor}
 
The isotropic form of the  fourth-order tensor 
$M_{ijkl} = \langle g_i g_j A_{kl} \rangle$
can be generally written as \cite{popebook}:
\begin{align}
    M_{ijkl} = \langle g_i g_j A_{kl} \rangle =\alpha\,\delta_{ij} \delta_{kl}
             + \beta\,\delta_{ik} \delta_{jl}
             + \gamma\,\delta_{il} \delta_{jk}.
    \label{eq:isotropic_form}
\end{align}
From the symmetry of $g_ig_j$ it follows that
$M_{ijkl} = M_{jikl}$, leading to $\beta = \gamma$; whereas
the incompressibility condition $\partial u_k/\partial x_k = 0$
leads to $M_{ijkk} = 0$, giving 
$3\alpha + \beta + \gamma = 0$. Combining the two relations
leads to
\begin{align}
    M_{ijkl} = \beta\!\left(
        -\frac{2}{3}\,\delta_{ij} \delta_{kl}
        + \delta_{ik} \delta_{jl}
        + \delta_{il} \delta_{jk}
    \right) \ . 
    \label{eq:Mijkl_beta}
\end{align}
Setting all indices to $1$ gives 
$M_{1111} = \langle g_1^2 A_{11} \rangle = \tfrac{4}{3} \beta$,
hence $\beta = \tfrac{3}{4} \langle g_1^2 A_{11} \rangle$, leading to
\begin{align}
    \langle g_i g_j A_{kl} \rangle
    = \frac{3}{4}\,\langle g_1^2 A_{11} \rangle
      \!\left(
        -\frac{2}{3}\,\delta_{ij} \delta_{kl}
        + \delta_{ik} \delta_{jl}
        + \delta_{il} \delta_{jk}
      \right) \ . 
\end{align}
which is the result given in Eq.~\ref{eq:Mijkl_final}
It is worth noting that while we have assumed the tensor
$M_{ijkl}$ to be isotropic, it is not strictly so in passive
scalar mixing at low $\re$ and $Sc$. The local isotropy
of the tensor is broken by the uniform mean scalar gradient.
This can be assessed by considering the longitudinal 
components of the tensor $\langle g_\alpha g_\alpha A_{\alpha \alpha} \rangle$,
for $\alpha = 1,2,3$, where repeated $\alpha$ does not imply summation.
Under perfect isotropy, one expects
\begin{align}
    \langle g_1^2 A_{11} \rangle
    = \langle g_2^2 A_{22} \rangle
    = \langle g_3^2 A_{33} \rangle
    = \frac{2}{15}\,\langle g_ig_jS_{ij} \rangle,
    \label{eq:diag_identity}
\end{align}
corresponding upon non-dimensionalization to the universal 
value $\mathcal{S}_{u\theta} \approx -0.5$ for each diagonal 
contribution \cite{Tang2025}.
However, as explored in previous works \cite{Yeung2002, DY2010, BCSY2021a},
at low $\re$ and $Sc$, 
the contribution parallel to imposed gradient
ends up being slightly larger than the contributions
in perpendicular directions.
With increasing $\re$ and $Sc$, the contributions
become equal and isotropy is recovered as the scalar gradient is compressed to increasingly smaller 
scales relative to the velocity field.

\clearpage

\section{Joint distributions corresponding to Fig.~\ref{fig:joint_pdfs}}
\label{app:joint_pdfs}

\begin{figure}[ht]
\centering
\includegraphics[width=0.9\textwidth]{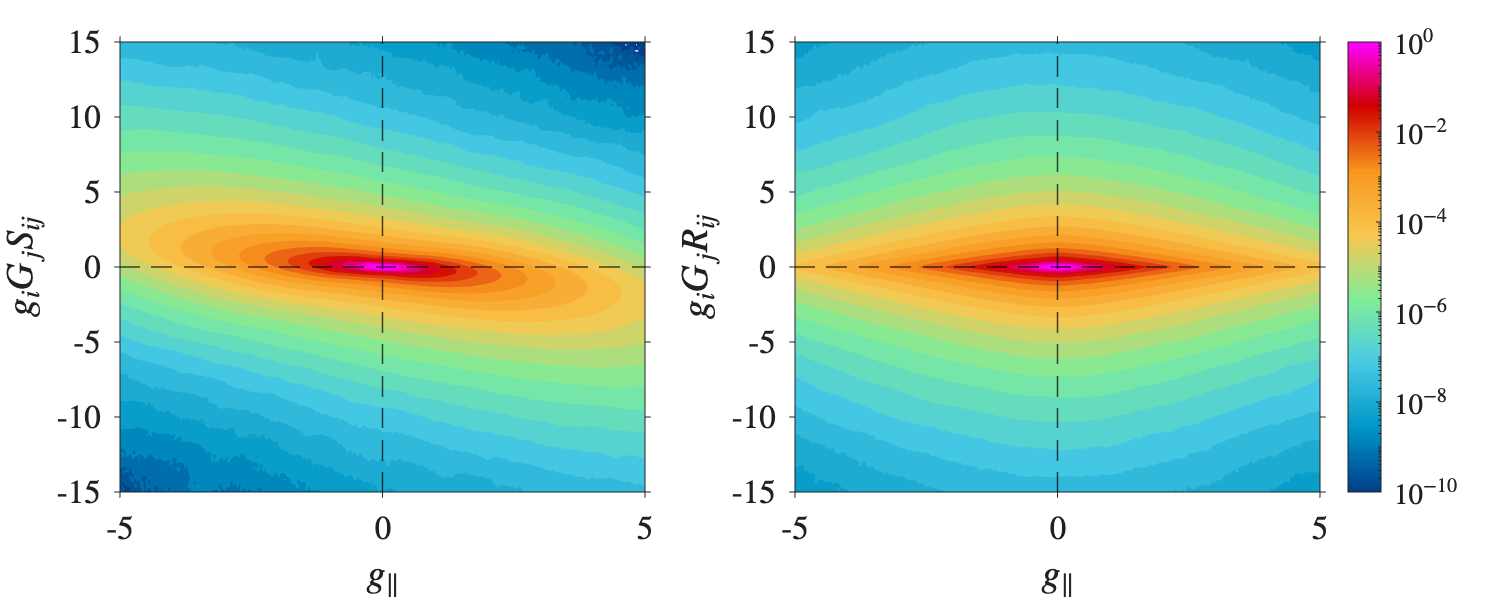}\\
\includegraphics[width=0.9\textwidth]{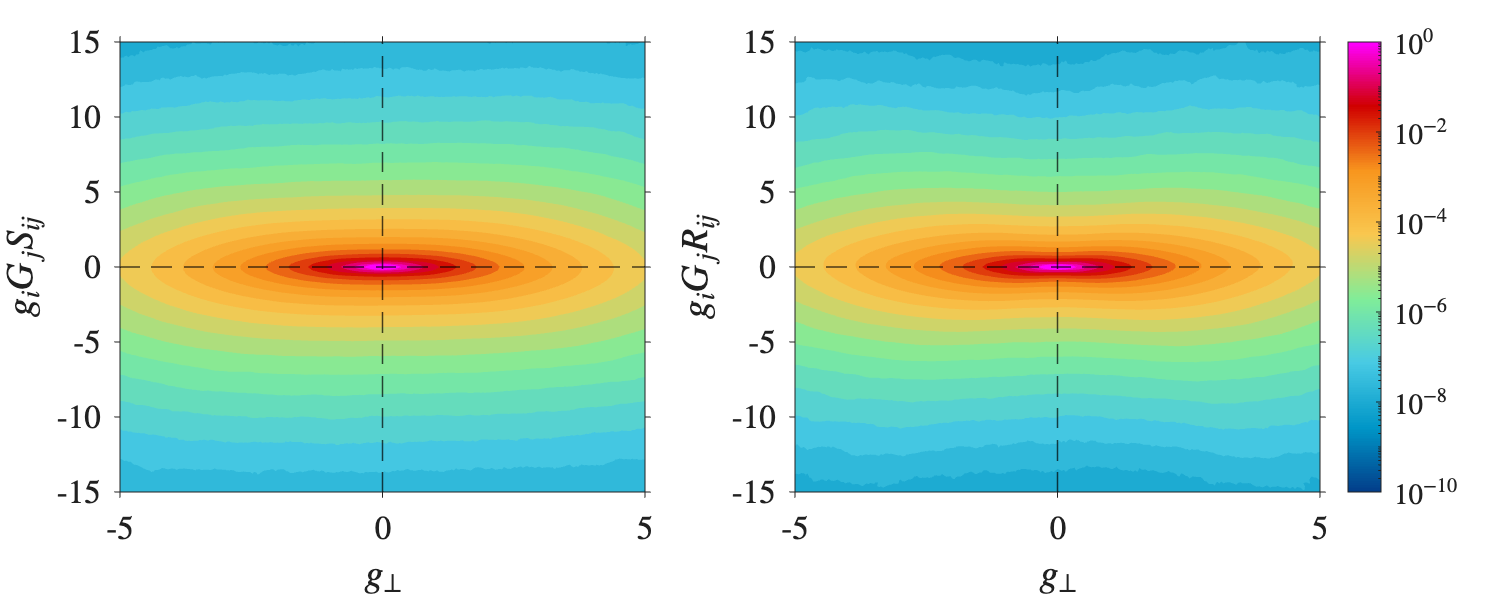}
\caption{
Joint probability density functions (p.d.f.s) of 
$g_\parallel$ (top row),  $g_\perp$ (bottom row), the 
scalar gradient components parallel and perpendicular to the 
imposed mean-gradient $\mathbf{G}$, with the amplification terms
$g_iG_jS_{ij}$ (left column), $g_iG_jR_{ij}$ (right column), 
at $Re_\lambda = 140$, $Sc = 512$.
The scalar gradients is normalized by its $L^2$-norm: $\langle g_ig_i\rangle^{1/2}$
and the strain and rotation tensors by the Kolmogorov time
scale $\tau_K$.
}
\label{fig:joint_pdf_140_512}
\end{figure}


%
\end{document}